\documentclass[reprint, aps, pra,noeprint]{revtex4-2}

\usepackage{amssymb, amsmath, amsfonts}
\usepackage[rgb]{xcolor}
\usepackage{silence}
\ActivateWarningFilters[latexPlacement]
\usepackage[hidelinks]{hyperref}
\usepackage{braket}
\usepackage{graphicx}
\usepackage[caption=false]{subfig}
\usepackage{microtype}

\DeclareMathOperator{\sinc}{sinc}
\DeclareMathOperator{\im}{Im}
\DeclareMathOperator{\Si}{Si}
\DeclareMathOperator{\si}{si}
\DeclareMathOperator{\Cin}{Cin}

\numberwithin{equation}{section}

\begin{document}

\title{Local quantum detection of cosmological expansion:\texorpdfstring{\\}{} Unruh-DeWitt in spatially compact Milne}
\author{Adam S. Wilkinson}
    \email{adam.wilkinson@nottingham.ac.uk}
\author{Jorma Louko}
    \email{jorma.louko@nottingham.ac.uk}
\affiliation{School of Mathematical Sciences, University of Nottingham, NG7 2RD, UK}
\date{July 2024, revised October 2024}

\begin{abstract}
    We analyse the excitations and de-excitations of an inertial Unruh-DeWitt detector in the $(1+1)$-dimensional expanding Milne cosmology with compact spatial sections, coupled to a real massless scalar field with either untwisted or twisted boundary conditions, prepared in the conformal vacuum. We find the detector's response as a function of the energy gap, the Milne spatial circumference parameter, the interaction duration, the age of the universe at the switch-on moment, the detector's peculiar velocity at the switch-on moment, and, for the untwisted field, the state of the zero mode. Asymptotic analytic results are obtained at large energy gap and at large circumference parameter, in each case recovering the Minkowski vacuum response in the leading order, and in the double limit of small circumference parameter and late cosmological time, recovering the response in a static Minkowski cylinder. Numerical results are given in the interpolating regimes. The results confirm the detector's sensitivity to both classical and quantum properties of its environment. 
\end{abstract}

\maketitle

\section{Introduction}

Quantum phenomena accessible to local observers, such as ourselves, are localised in space and time. 
In relativistic quantum field theory, however, the quantum state of the field may encode nonlocal information, 
a prime example being the Reeh-Schlieder correlations in Minkowski vacuum \cite{Reeh1961}. 
In particular, in an expanding cosmology, the state of a quantum field would typically be set by initial conditions in the early universe 
\cite{Birrell1982,Mukhanov2007}, 
and these initial conditions may be sensitive to the spatial topology \cite{Levin2001}.
Local quantum observations may therefore carry an imprint of the history and the global properties of the spacetime \cite{Ford1989,Toussaint2021,Garay2014,Toussaint2024}.

In this paper we consider a local quantum observer in an expanding Milne universe in $1+1$ spacetime dimensions. 
The Milne universe is a Friedmann-Lema\^{i}tre-Robertson-Walker 
(FLRW) cosmological spacetime whose scale factor expands at constant rate,
it has vanishing spacetime curvature, and in four spacetime dimensions it has negative spatial curvature \cite{Milne1932nature,Rindler:2006km}. 
While a constant expansion rate has limited applicability in the description of own universe \cite{Nielsen2016,Avelino2016,Tutusaus2017,Casado2020,Mohayaee2021}, 
we shall utilise the Milne universe as an arena for analysing local observations of a globally-defined quantum state in a mathematically simple setting. 
In particular, the absence of spatial curvature in the $(1+1)$-dimensional Milne spacetime allows us to disentangle effects due to the size of the universe from those due to the universe's age and expansion rate. 

Concretely, we consider a $(1+1)$ Milne spacetime with spatial topology~$S^1$. 
In this spacetime, the ratio of the spatial circumference and the cosmological time remains constant as the universe evolves, 
and the value of the constant is a freely-specifiable ``universe size'' parameter. 
Classically, the universe size parameter can be determined by experiments involving communication that circumnavigates the universe, 
for example by sending and recapturing light rays. 
Our interest is to observe the universe size parameter via local quantum equipment, 
through the imprint of this parameter in the conformal vacuum of a quantised real massless scalar field. 

As the local quantum equipment, we furnish an observer with an Unruh-DeWitt detector \cite{Unruh1976,DeWitt1979}, a spatially pointlike two-state quantum system, linearly coupled to the field. The detector is characterised only by the gap between its two energy levels, and it can be regarded as a simplified model of atomic interactions with the electromagnetic field \cite{Martinez2013,Alhambra2014}. 
To avoid effects due to acceleration, we consider observers who are inertial but may have a nonzero peculiar velocity, 
defined as the velocity with respect to the co-moving cosmological observers. 

To summarise, the classical input consists of the universe size parameter, the age of the universe when the detector is switched on, the detector's peculiar velocity at the switch-on event, and the duration of the interaction in the detector's proper time. 
The quantum input consists of the detector's energy gap, 
the choice of the conformal vacuum as the field's initial state, 
and 
a potential subtlety 
concerning zero modes, described in the next paragraph.  
The output consists of the detector's response, defined as the transition probability between the detector's two states. 
We shall analyse how the detector's response depends on the classical and quantum input, including the universe size parameter. 

The remaining potential subtlety is that a real scalar field on a spacetime with $S^1$ spatial topology can have either periodic or antiperiodic boundary conditions 
in the spatial~$S^1$. 
For periodic boundary conditions, a massless field has a zero-momentum field mode, known as the zero mode, and this mode does not admit a conformal Fock vacuum \cite{Birrell1982,Rajaraman1982}. 
Additional quantum input is hence required to specify the zero mode state \cite{Allen1987,Ford1989,Garriga1992,McCartor1992,Kirsten1993}, 
and this additional input affects the detector's response \cite{Martinez2014,Toussaint2021,Louko2016,Tjoa2020}. 
A similar subtlety arises even for massive fields in cosmologies with certain types of initial singularities \cite{Ford1989,Toussaint2021}, 
and the detector's response for a massive scalar field in $1+1$ Milne under this subtlety was addressed in~\cite{Toussaint2021}. 
Our setting is an extension of \cite{Toussaint2021} to a massless field, exploring a complementary region of the parameter space, 
and using special features of the massless field to corroborate numerical simulations with analytic results in asymptotic regimes. 

One general feature evident in our results is that excitation probabilities 
are lower than de-excitation probabilities, and this disparity grows as the energy gap increases, as one would expect. 
The detailed profile of the transition probabilities as a function of the energy gap exhibits additional smaller scale structure, 
in the form of resonance peaks whose locations depend on the other parameters; in particular, the distance between the peaks decreases when the interaction duration increases. 
Increasing the duration also increases the overall magnitude of the probability, and for the antiperiodic field this increase is approximately linear, 
as one might expect, whereas the increase for the periodic field is more complicated due to significant contributions from the zero mode.

Varying the universe size parameter is more subtle.
With the other parameters fixed, in the limit of a large size parameter the response develops an infrared divergence, 
characteristic of a massless field in $1+1$ dimensions, whereas in the limit of a small size parameter the response decays to zero, 
due to the increasing frequency of the lowest positive frequency field mode. 
However, if the limit of a small size parameter is taken simultaneously with the limit of late cosmological time, 
keeping the product of the two constant, we recover the response to a field on a static cylinder~\cite{Martinez2014}. 

The detector's peculiar velocity has only a minor effect on the locations of the peaks in the response, 
but the heights of the individual peaks increase significantly as the peculiar velocity increases, 
and this effect is greatest in the peaks at small values of the energy gap. 
The large gap behaviour, by contrast, is close to the large gap behaviour in Minkowski vacuum, where a peculiar velocity is not defined.

For the periodic field, the contribution to the detector's response from the zero mode introduces additional excitation and de-excitation structure 
at small values of the energy gap, and this structure is sensitive to the choice of the zero mode state, the interaction duration, and the universe size parameter, 
but it is less sensitive to the detector's peculiar velocity. 

The structure of the paper is as follows. 
In section \ref{sec:thespacetime} we introduce the spatially compactified $1+1$ Milne spacetime, 
and we describe how an inertial observer in this spacetime can perform a classical experiment, 
sending light rays around the universe, 
to determine the size and age of the universe and their own peculiar velocity, 
despite the geometry being locally Minkowski. 
Section \ref{sec:qft} presents first the Fock quantisation of a real massless scalar field in this spacetime, 
with both periodic and antiperiodic boundary conditions, 
and then sets up the formulas for the response of an inertial Unruh-DeWitt detector that is linearly coupled to this field. 
The main results of the paper consist of the analysis of the detector's response in sections \ref{sec:analyticlimits} and \ref{sec:numerics}, 
in section \ref{sec:analyticlimits} by analytic methods in several asymptotic regimes, and in section \ref{sec:numerics} by numerical simulations. 
Section \ref{sec:discussion} gives a brief summary and discussion. 
The figures produced by the numerical analysis of section \ref{sec:numerics} are collected in appendix~\ref{app:figures}, 
and the asymptotic analysis of the large gap regime is collected in appendices \ref{app:largePositiveMu}, \ref{app:largeNegativeMu}, and~\ref{app:minkAsym}. 

We use the sign convention $(-,+)$, such that timelike squared distances are negative, and units in which $c=\hbar=1$. 

\section{The Spacetime Geometry\label{sec:thespacetime}}

\subsection{Milne Metric}
\label{subsec:milnemetric}

The Milne spacetime is a simple expanding cosmology which dates back to the 1930s \cite{Milne1936,Walker1937} whose most notable feature is a constant rate of expansion, otherwise known as a ``coasting'' model \cite{Casado2020}. Such a model is clearly at odds with the conventional $\Lambda$CDM model which predicts an accelerating rate of expansion, but there is some debate whether current observation rules out such a model \cite{Nielsen2016,Rubin2016}, most recently with the questioning of assumptions made when analysing supernovae Ia data \cite{Tutusaus2017,Mohayaee2021}. A comprehensive history of the cosmological context can be found in \cite{Casado2020}.

The Milne cosmology also has several enticing features when compared with $\Lambda$CDM models, not requiring any inflationary theory to explain the flatness fine-tuning problem or the horizon problem \cite{Batra2000,Gehlaut2003,Casado2020} and solving other issues such as the synchronicity problem \cite{Avelino2016}.

\begin{figure}[ht]
    \centering
    \includegraphics{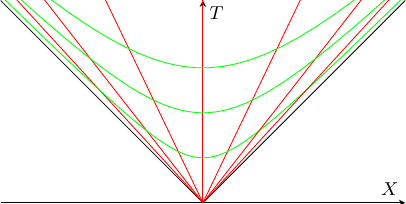}
    \caption{$1+1$ Milne spacetime shown as the future quadrant $|X| < T$ in $1+1$ Minkowski space with the Minkowski coordinates $(T,X)$. 
    In terms of the coordinates $(t,x)$ in~\eqref{eq:milneMetric}, 
    the (green) hyperbolas are spacelike surfaces of constant~$t$,  
    and the straight (red) lines are timelike surfaces of constant~$x$. 
The coordinate singularity $t=0$ consists of the future light cone $T = |X|$.}
    \label{fig:inertialPath}
\end{figure}

The Milne model is used here as a basis for investigating quantum mechanics in expanding spacetimes in a similar context to previous work~\cite{Toussaint2021}. 
It can be obtained from the future wedge $|X| < T$ in Minkowski space with the Minkowski coordinates $(T,X)$, by the coordinate transformation 
\begin{subequations}
\begin{align}
T &= t\cosh(ax),
\\
X &= t\sinh(ax), 
\end{align}
\end{subequations}
where $0<t<\infty$, $-\infty < x < \infty$, and $a$ is a positive parameter of dimension inverse length. 
The resulting metric is
\begin{align}
    \mathrm{d}s^2 = -\mathrm{d}t^2 + a^2t^2\mathrm{d}x^2 . 
    \label{eq:milneMetric}
\end{align}
Figure \ref{fig:inertialPath} shows a spacetime diagram with the old and new coordinates. 

As the Milne universe is just a coordinate transformation in a subset of Minkowski spacetime, no geometric experiment 
will be able to distinguish between the two. 
To obtain a $(1+1)$-dimensional expanding spacetime with an invariant notion of spatial size, 
we introduce a spatial periodicity by the identification $(t, x) \sim (t, x+L)$, where $L$ is a positive parameter of dimension length. 
In terms of the Minkowski geometry, this means introducing a periodicity under a boost of rapidity~$aL$, and therefore the outcomes of any experiments should depend on $a$ and $L$ only through the dimensionless combination~$\xi := aL$, which we call 
the ``universe size parameter''.
The periodic identification makes the former coordinate singularity at $T = |X|$ a true singularity \cite{Hawking1973}, and Poincaré invariance is broken in the system as is true with any quotient of Minkowski \cite{Uzan2002}. 
Specifically, inertial observers possess an invariant notion of a peculiar velocity, as we shall discuss in subsection \ref{subseq:classical-experiment} below.

\subsection{A Classical Experiment} 
\label{subseq:classical-experiment}

We now describe a classical experiment by which an inertial observer in our spatially compact Milne universe can determine the universe size parameter~$\xi$, 
the age of the universe, and their peculiar velocity. This experiment will require information to travel around the spacetime, and it would therefore not be feasible for cosmologically large universes. Similar experiments can be carried out in other geometries that are quotients under a discrete isometry, 
such as Minkowski spacetime identified under a spatial translation, and they can be used to clarify apparent paradoxes in relativistic communication scenarios \cite{Dray1990,Low1990,Uzan2002}.  

Recall that an observer's peculiar velocity in a FLRW metric is defined as the velocity relative to the co-moving reference frame at the same event~\cite{Weinberg2013}. 
In an expanding FLRW metric, the peculiar velocity of any non-comoving inertial observer will hence decrease in time. 
For an inertial observer moving through our spatially compact Milne~\eqref{eq:milneMetric}, 
an experiment should therefore be able to determine not only the universe size parameter~$\xi$, 
but also the age of the universe $t_0$ and the observer's peculiar velocity $v$ at the event where the observer performs their experiment. 

\begin{figure}[t]
    \centering
    \includegraphics{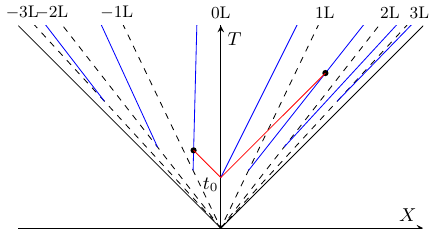}
\caption{A spacetime diagram of a classical experiment by which an inertial observer can determine the universe size parameter~$\xi$,  
their peculiar velocity, and the cosmological time. 
The straight (blue) lines are multiple copies of the observer's worldline, embedded in the future quadrant of $1+1$ Minkowski, and identified by $(t, x) \sim (t, x+L)$.
A pair of light beams (red), at 45 degrees from the vertical, are shown emitted by one copy of the observer at cosmological time $t_0$ and recaptured by neighbouring copies of the observer at the black dots, having circumnavigated the universe. The observer in the figure has a positive peculiar velocity, and hence the left-moving light beam is captured earlier than the right-moving light beam.}
    \label{fig:milneClassicalExperiment}
\end{figure}

An experiment that accomplishes this is as follows. 
The observer shines a light beam simultaneously in both the right (increasing $x$) and left (decreasing $x$) 
directions at cosmological time~$t_0$, starting a stopwatch. 
The observer records their proper time $\tau_R$ when the right-moving light beam returns 
and their proper time $\tau_L$ when the left-moving light beam returns. 
A spacetime diagram of the experiment is shown in figure~\ref{fig:milneClassicalExperiment}. 
The observer also measures~$\tilde{f}$, the ratio of the frequency emitted over the frequency received. 
In terms of the underlying Minkowski geometry, $\tilde{f}$ can be understood as the Doppler shift 
between two inertial observers receding from each other with rapidity~$\xi$, 
which means that the light will be redshifted, and $\tilde{f} = e^{\xi}>1$. 
In terms of the observed quantities $(\tau_R,\tau_L,\tilde{f})$, we then find 
\begin{subequations}
    \begin{align}
        \xi &= \ln\tilde{f} , \\
        v &= \frac{\tau_R-\tau_L}{\tau_R+\tau_L} ,
        \label{eq:v-measurement}
        \\
        t_0 &= \frac{\sqrt{\tau_R\tau_L}}{\tilde{f}-1} , 
    \end{align}
\end{subequations}
where $v$ is the peculiar velocity at the event where the light beams are sent. 
Note from \eqref{eq:v-measurement} that the observer is comoving if and only if the right-moving and left-moving beams are received at the same moment.

This classical experiment illustrates the accessibility of the system parameters to a classical observer, 
but the experiment is necessarily global in that it requires a circumnavigation of the universe. We next consider a quantum experiment that is local, and is thus better suited to a large universe such as our own.

\section{Quantum Field Theory \label{sec:qft}}
\subsection{Constructing the Field}
For writing down the quantum field theory, it is convenient to express the metric \eqref{eq:milneMetric} in conformal coordinates $(\eta, x)$ where 
\begin{align}
    \eta &= a^{-1}\ln(at)
\end{align}
is the conformal time. The metric then takes the form
\begin{align}
    \mathrm{d}s^2 = e^{2a\eta}\left(-\mathrm{d}\eta^2 + \mathrm{d}x^2\right)
\end{align}
therefore giving solutions that are conformal to a slab of Minkowski. In these conformal coordinates, we then work with the action for a real, massless, scalar Klein-Gordon field
\begin{align}
    S = -\frac{1}{2}\int \sqrt{-g}\,\mathrm{d}\eta\,\mathrm{d}x\, \nabla_\mu\phi\nabla^\mu\phi
\end{align}
which gives the field equation
\begin{align}
    (-\partial_\eta^2 + \partial_x^2)\phi = 0.
\end{align}
At this point, the compact spacetime presents us with two valid boundary conditions under the quotient map which results in two distinct field theory solutions. Perhaps the more intuitive option is the untwisted case, which has periodic boundary conditions such that the quotient map $g$ transforms the field as $g: \phi\to\phi$. A twisted field solution is also possible, which transforms as $g:\phi\to-\phi$ giving anti-periodic boundary conditions. It is worth noting here that for a complex-valued field any phase change is possible under the quotient map~\cite{Higuchi:2022nfy}, but since we restrict ourselves to a real-valued field only sign changes are possible.

At this point, we are equipped to solve for mode solutions of the field, normalised under the Klein-Gordon inner product
\begin{align}
    (U_n,U_m) &= i\int \mathrm{d}x (U_n^*\partial_\eta U_m - U_m\partial_\eta U^*_n) = \delta_{nm}.
\end{align}
Conformal coordinates allow us to find the mode function solutions with positive conformal frequency as in~\cite{Martinez2014}, with the result  
\begin{align}
    U_n(x, \eta) &= \frac{1}{\sqrt{2\left|k_n\right|L}}\exp\left(-i\left|k_n\right|\eta + ik_n x\right) , 
\end{align}
where $L$ is the period of $x$ as introduced in section~\ref{subsec:milnemetric}, 
and $k_n$ varies depending on the twisted or untwisted nature of the field, 
\begin{align}
    k_n &= \begin{cases}
        2n\pi/L &    \mathrm{Untwisted},n\neq 0, \\
        (2n+1)\pi/L & \mathrm{Twisted.}
    \end{cases}
\end{align}
Note here that the untwisted field's oscillatory solution is divergent for $n=0$. This field mode with zero frequency requires separate analysis, solving the field equation for a mode which is constant in $x$ we find the normalised zero mode of the field as
\begin{align}
    U_{zm} &= \frac{\eta + \zeta}{\sqrt{2L \im(\zeta)}} , 
\label{eq:Uzm-def}
\end{align}
where $\zeta$ a complex parameter satisfying $\im(\zeta) > 0$. It is worth noting that if we write this zero mode in coordinate time, it takes the form
\begin{align}
    U_{zm}(t, x) &= \frac{\ln(at) + a\zeta}{\sqrt{2aL \im(a\zeta)}} , 
\end{align}
such that the dimensionless parameter combination $a\zeta$ presents itself as simpler representation of this zero mode ambiguity, and the geometric parameters combine into $aL$ as is expected given the geometric arguments in section \ref{sec:thespacetime}.

With this full set of field modes, noting this distinct non-oscillatory behaviour of the $n=0$ mode of the untwisted field, we can 
expand the quantum field as 
\begin{align}
    \hat{\phi}(x, \eta) &= \sum_n U_n(x, \eta)a_n + U^*_n(x, \eta)a_n^\dagger , \label{eq:fieldop} 
\end{align}
where the creation and annihilation operators satisfy the usual commutation relations 
\begin{align}
    [a_n,a_m^\dagger] = \delta_{n,m} . 
\end{align} 
It follows that $\phi$ satisfies the canonical commutation relations. A vacuum $\ket{0}$ is then defined as the state annihilated by all the annihilation operators, 
\begin{align}
    a_n\ket{0}=0\hspace{1em}\forall n . 
    \label{eq:conformalVacuum}
\end{align}
The vacuum defined in this way is called the conformal vacuum with respect to the oscillatory modes of the field since it is defined with respect to the annihilation operators of the conformal field modes, and is the conventional choice for the field state before any interactions. Since the non-oscillatory zero mode has a parametric freedom, a unique preferred vacuum state cannot be defined, and each value of $\zeta$ will correspond to a different vacuum state. The impact of this $\zeta$ parameter in the massive case is analysed at length in \cite{Toussaint2021}, and its impact on an observer's measurements in our massless case will be investigated numerically in section \ref{subsec:zero mode numerics}.

\subsection{Detector\label{subsec:detector}}
To probe the quantum field established above, we utilise an Unruh-DeWitt detector \cite{Unruh1976,DeWitt1979}. This is a qubit coupled to the field in a way analogous to the electromagnetic field interaction with atomic orbitals, being a good leading-order approximation in processes without angular momentum exchange \cite{Martinez2013,Alhambra2014}. To begin, first establish the qubit subsystem as having two accessible states, a state with zero energy, $\ket{0}_D$, and second state with energy $\omega$, $\ket{\omega}_D$. Note that we place no restrictions on the sign of~$\omega$, a freedom that will be clarified below.

We then couple this system to the field via the interaction Hamiltonian
\begin{align}
    H(\tau) := \lambda\,\chi(\tau)\,\mu(\tau)\,\phi(\mathsf{x}(\tau)) , \label{eq:interaction-hamiltonian}
\end{align}
where $\lambda$ is a coupling constant, $\chi(\tau)$ is a real-valued switching function that controls the strength of the coupling, $\mu(\tau)$ is the monopole moment operator of the detector, 
\begin{align}
    \mu(\tau) &= \mathrm{e}^{i\omega\tau}\ket{\omega}_D\bra{0}_D + \mathrm{e}^{-i\omega\tau}\ket{0}_D\bra{\omega}_D , 
\end{align} 
and $\phi(\mathsf{x}(\tau))$ is the field operator to which the detector is coupled, pulled back to the detector worldline $\mathsf{x}(\tau)$.

We prepare the system so that before the interaction the detector is in its zero energy state $\ket{0}_D$ and the 
field is in the state $\ket{0}$ \eqref{eq:conformalVacuum}, 
which is the conformal vacuum for the oscillator modes, 
and for the zero mode of the untwisted field 
it is the state pararametrised by the zero-mode parameter $\zeta$ in~\eqref{eq:Uzm-def}.
Expanding the interaction \eqref{eq:interaction-hamiltonian} in time-dependent perturbation theory in the coupling constant $\lambda$, we find that the leading order probability of the detector transitioning to the state~$\ket{\omega}_D$, regardless the final state of the field, is proportional to the response function \cite{Birrell1982,Junker:2001gx}
\begin{align}
    \mathcal{F}(\omega) &= \int_{-\infty}^\infty\int_{-\infty}^\infty\mathrm{d}\tau\mathrm{d}\tau' \chi(\tau)\chi(\tau')\mathrm{e}^{-i\omega(\tau-\tau')}G(\tau, \tau') , 
\end{align}
where
\begin{align}
    G(\tau, \tau') = \bra{0}\phi(\mathsf{x}(\tau))\phi(\mathsf{x}(\tau'))\ket{0}
\end{align}
is the pull-back of the Wightman function of the field to the detector's worldline. The detector's transition is an excitation for $\omega>0$ and a de-excitation for $\omega<0$. 

Next, we address the choice of $\chi$. We will use a sharp switching function, with an instantaneous switch-on at time $\tau_i$ and switch-off at time $\tau_f$, such that
\begin{align}
    \chi(\tau) := \Theta(\tau-\tau_i)\Theta(\tau_f - \tau).
\label{eq:switching-choice}
\end{align}
This choice of switching allows us to reduce the number of variable parameters in the system, since a smoother switching function would introduce further parametric dependence in the final response function. There will still be some effect from the switching itself, but that effect will be constant. A further benefit is the simplification of the response function to
\begin{align}
    \mathcal{F}(\omega) &= \int_{\tau_i}^{\tau_f}\int_{\tau_i}^{\tau_f}\mathrm{d}\tau\mathrm{d}\tau'\mathrm{e}^{-i\omega(\tau-\tau')}G(\tau, \tau').
\end{align}
This sudden switching produces irregularities and non-lorentzian contributions in dimensions greater than (2+1), which are often addressed by introducing an infinitesimal spatial extent \cite{Schlicht2004,Louko2007}, or by defining the switching function as a limiting case \cite{Satz2007}, and it is impossible to regularise in more than 5 dimensions \cite{Hodgkinson2012}.

We further specialise our system to only look at inertial observers. We define the velocity parameter $v$ as the detector's peculiar velocity at $(t,x) = (t_0,0)$, identically to the classical experiment shown in figure \ref{fig:milneClassicalExperiment}. Without loss of generality, we also take the proper time of the observer as zero at this initial point $t_0$, such that times afterwards are effectively durations from that point and $\tau_f$ is the duration of the interaction.

Before writing the transition rate for our system, let us first identity our system parameters. Just defined is the observer velocity $v$, but other dimensionless parameters include the universe size parameter $\xi=aL$, and the zero mode ambiguity which we will write as two real parameters such that $a\zeta=q\mathrm{e}^{i\theta}$ and $0 < \theta < \pi$, $q > 0$.

There are also three dimensionful parameters present in our system, the cosmological time at switch on $t_0$, the interaction duration $\tau_f$, and the detector gap parameter $\omega$. To simplify and de-dimensionalise these parameters, we choose $t_0$ as a convenient reference scale which we use to define
\begin{subequations}
    \begin{align}
        \mu &= \omega t_0,\\
        \sigma &= \tau / t_0 , 
    \end{align}
\end{subequations}
as our dimensionless gap parameter $\mu$ and our dimensionless proper time $\sigma$, such that $\sigma_f = \tau_f/t_0$ is the dimensionless interaction duration. In terms of these parameters, the response function of the detector to the oscillatory field modes can be written
\begin{align}
    \mathcal{F}_{osc} &= \frac{t_0^2}{2} \sum_{n=1}^\infty \frac{1}{M_n}\left(\left|f_n^+\left(\mu\right)\right|^2 + \left|f_n^-\left(\mu\right)\right|^2\right) , \label{eq:detectorResponse}
\end{align}
where
\begin{subequations}
    \begin{align}
        f_n^\pm(\mu) &= \int_{0}^{\sigma_f} \mathrm{d}x\; \mathrm{exp}\left[ i\left( \mu x + \frac{M_n}{\xi}\ln(x+\alpha_\pm) \right) \right] , \\
        \alpha_\pm &= \sqrt\frac{1\pm v}{1\mp v} , 
    \end{align}
\end{subequations}
are the contributions from left and right moving mode functions, and $M_n$ differs for twisted and untwisted fields as
\begin{align}
    M_n &= 
    \begin{cases}
        2n\pi & \mathrm{Untwisted,} \\
        (2n-1)\pi & \mathrm{Twisted.}
    \end{cases}  \label{eq:Mn}
\end{align}
The untwisted field's zero mode gives an additional response of the form
\begin{align}
    \mathcal{F}_{zm} &= t_0^2\frac{\left| g^+\left(\mu\right) + g^-\left(\mu\right) - 2q\mathrm{e}^{i\theta}\left(1-\mathrm{e}^{-i\mu\sigma_f}\right) \right|^2}{{8\xi q\sin\theta\mu^2}} , 
\label{eq:zeroModeResponse}
\end{align} 
where the functions $g^\pm$ are given by
\begin{align}
    g^\pm(\mu) &= \mathrm{e}^{-i\mu\sigma_f}\ln\left(\sigma_f + \alpha^\pm\right) \notag\\&\hspace{2em}- \mathrm{e}^{-i\mu\alpha^\pm}\biggl[ E_1\left(i\mu\alpha^\pm\right) - E_1(i\mu(\sigma_f+\alpha^\pm)) \biggr] , 
\end{align}
and $E_1$ is the principal branch of the complex analytic continuation of the exponential integral as defined in \cite[~Eq. 6.2.1]{DLMF}. 

We now take several analytical limits and turn to numerical modelling in the interpolating regimes to understand the sensitivity of the detector to system parameters.

\section{Asymptotic Regimes\label{sec:analyticlimits}}

This section comprises a discussion of several analytical limiting cases of the detector's response function. We investigate the impact of large detector energy gap and note the differing behaviour of excitations and de-excitations, followed by limits of large and small universe size parameter~$\xi$, and the simultaneous small size, late time limit.

\subsection{Large Energy Gap\label{subsec:largeSmallMu}}

In this subsection we find asymptotic expansions of the detector's response function \eqref{eq:detectorResponse} for large positive and large negative gap values $\omega$. Large positive gap corresponds to a large excitation gap, requiring a lot of energy to change the detector state. Conversely, large negative gap corresponds to a large de-excitation where the detector loses a large amount of energy, so it is expected to be more likely. Detailed calculations are shown in appendices \ref{app:largePositiveMu}, \ref{app:largeNegativeMu} and show the large excitation gap behaviour to follow
\begin{align}
    \mathcal{F}(\omega) &= \frac{\ln(t_0\omega)}{\pi\omega^2} + \mathcal{O}(\omega^{-2}) , \label{eq:largePositiveMu}
\end{align}
and large de-excitation gap to follow
\begin{align}
    \mathcal{F}(\omega) &= \frac{\tau_f}{|\omega|} + \mathcal{O} \bigl(\omega^{-2}\ln|\omega|\bigr) , \label{eq:largeNegativeMu}
\end{align}
in line with expectations that de-excitation is more likely. It is found that the zero mode of the untwisted field is sub-leading in the detector response for large $\omega$, and the behaviour of both twisted and untwisted fields is identical to leading order. 

Note that the error term in \eqref{eq:largeNegativeMu} should be understood as $\mathcal{O} \bigl( \omega^{-2}\ln|\omega/\omega_0| \bigr)$, where $\omega_0$ is an arbitrary positive constant of dimension inverse length, making the expression dimensionally consistent. The value of $\omega_0$ is of no consequence, since changing $\omega_0$ adds to $\omega^{-2} \ln|\omega/\omega_0|$ a multiple of $\omega^{-2}$, which is subdominant. We use a similar shorthand in error terms involving logarithms in the rest of the paper. 

\begin{figure}[ht]
    \centering
    \includegraphics[width=0.48\textwidth]{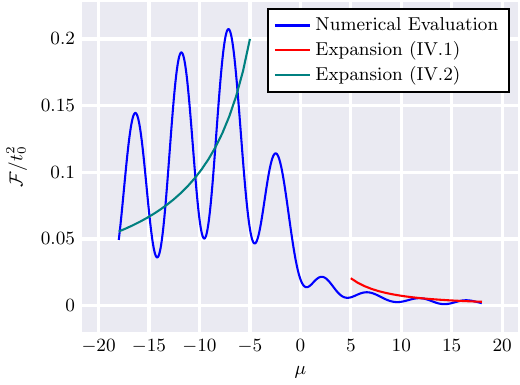}
    \caption{The response function for the twisted field as a function of the dimensionless energy $\mu = \omega t_0$, with the parameter values $\xi = 10$, $\tau_f/t_0 = 1$ and $v=0$ (blue). The graph shows oscillations of characteristic period of approximately $5$, with a falloff at large $|\mu|$ and a suppression of excitations compared with de-excitations, as was to be expected. The falloff at large positive $\mu$ is consistent with the leading term in the asymptotic expansion \eqref{eq:largePositiveMu}, shown in red, but containing still oscillations around this term within the plotted range. Similar behaviour is seen for the falloff at large negative $\mu$ and the leading term in the asymptotic expansion \eqref{eq:largeNegativeMu}, shown in green.}
    \label{fig:asymComparison}
\end{figure}

Figure \ref{fig:asymComparison} shows these expansions alongside the numerically evaluated field response, and there is good correspondence especially for large $\omega$.

These expansions also show identical leading order behaviour to the large $\omega$ expansion in infinite Minkowski, not only in the dependence on $\omega$, but also the coefficient of the leading response. This quantitative agreement, detailed in appendix \ref{app:minkAsym}, shows that as the gap gets larger the system is less sensitive to the universe size, a proposition that is reinforced by the numerics in section \ref{sec:numerics}.

\subsection{Small and Large Spatial Sizes\label{subsec:spatialSizeLimits}}

We here investigate the impact of small and asymptotically large values of the universe size parameter~$\xi$. 

\subsubsection{Oscillator modes}

The response of the detector to the oscillatory field modes shows similar asymptotic behaviour to periodic Minkowski, otherwise called the static cylinder universe. 
As the universe size approaches zero, the response of the detector to the oscillatory modes of the field \eqref{eq:detectorResponse} vanishes. 
This is behaviour that can be understood in terms of the energy of the oscillatory modes, since their minimum energy increases as the universe size decreases. 
As their minimum energy increases, the detector will be less able to exchange energy with these modes, and therefore its response to them will decrease. 
As the size of the universe approaches zero, their energy diverges and they invoke no response in the detector.

For large values of the universe size parameter, an infrared divergence appears in the detector response to the oscillatory modes \eqref{eq:detectorResponse}. This is symptomatic of the infrared divergence present in the (1+1)-dimensional Minkowski Wightman function for a massless field, which often requires separate careful treatment by the introduction of an infrared cut-off \cite{Louko2014}.

As an alternative to these universe size limits, we also note here a simultaneous transformation which can be used to recover the response of the detector to a field defined on the static cylinder. We present a simultaneous late time and small universe size limit via direct comparison to periodic Minkowski with the spatial extent of the Milne universe at the switch-on time $t_0$, such that
\begin{align}
    l = t_0\xi.
\end{align}
With this comparison, it is possible to take asymptotic limits of $\xi$ while maintaining a fixed finite cylinder size $l$ by moving $t_0$ along with it, such that $\xi\to0$ and $t_0\to\infty$. To see this explicitly, we first recall the response of the detector only to the oscillatory field modes, postponing investigation of the zero mode for the time being. We write the Milne response \eqref{eq:detectorResponse} as $\mathcal{F}_\mathcal{M} = \mathcal{F}_\mathcal{M}^+ + \mathcal{F}_\mathcal{M}^-$, where
\begin{align}
    \mathcal{F}_\mathcal{M}^\pm(\omega) &= \frac{1}{2}\sum_{n=1}^\infty\frac{1}{M_n}\Biggl| \int_0^{\tau_f}\mathrm{d}\tau\notag\\&\hspace{2em}\left.\exp\left[ i\left( \omega\tau + M_n\ln\left[\left(1+\frac{\alpha_\pm\tau}{t_0}\right)^\frac{1}{\xi}\right] \right) \right] ^{\vphantom{\frac{a}{a}}}\right|^2
\end{align}
represent the contributions from left-moving and right-moving mode functions. In this form, replacing $t_0$ with $l/\xi$ recovers a finite limiting case. As $\xi\to0$ the limit creates a natural logarithm of an exponential, simplifying to
\begin{align}
    \mathcal{F}_\mathcal{M}^\pm &\to \frac{1}{2}\sum_{n=1}^\infty\frac{1}{M_n}\left|\int_0^{\tau_f}\mathrm{d}\tau\;\exp\left[i\tau\left(\omega + \frac{M_n\alpha_\pm}{l}\right)\right]^{\vphantom{\frac{a}{a}}}\right|^2. \label{eq:cylinderResponse}
\end{align}
This can be verified to be the response of an inertial detector in a static cylinder by comparison to the results of \cite{Martinez2014}.

\subsubsection{Zero mode}

Analysis of the detector's response to the zero mode of the untwisted field yields further insights, distinct to that of the oscillatory modes. For a fixed value of the zero mode ambiguity parameter $\zeta$, as $\xi\to0$ the detector's response to the zero mode diverges where its response to the oscillatory mode vanishes, as is evident from the $1/\xi$ dependence in the detector's response \eqref{eq:zeroModeResponse}. Thus, the detector's response to a small universe will be dominated by the zero mode response.

As the size parameter of our periodic Milne geometry is increased, $\xi\to\infty$, the response of the detector to the zero mode \eqref{eq:zeroModeResponse} tends towards zero. This is in line with expectations from the oscillatory modes since $\xi\to\infty$ again recovers Minkowski behaviour, wherein there is no zero mode.

As another limiting case, we again take the simultaneous late time, small universe limit. However, our zero mode parameter $q$ is made dimensionless by comparison to our geometric Milne parameters, and therefore needs to be appropriately transformed. Given the definition of the zero mode parameters from section \ref{subsec:detector}, 
\begin{align}
    a\zeta = q\mathrm{e}^{i\theta} , 
\end{align}
we transform $q$ as
\begin{align}
    \zeta=\frac{q}{a}\mathrm{e}^{i\theta} = \frac{lq}{\xi}\mathrm{e}^{i\theta} = lp\mathrm{e}^{i\theta} , 
\end{align}
such that $p=q/\xi$. With this transformation, we take the $\xi\to0$, $t_0\to\infty$ limit in the zero mode after mode function substitution but before simplifying the integral into known functions. This makes it easier to show that 
\begin{align}
    \mathcal{F}_{zm}(\omega) \to \frac{1}{2p\sin\theta}\left|\int_0^{\tau_f}\mathrm{e}^{i\tau\omega}\left( \frac{\tau}{\gamma} + lp\mathrm{e}^{i\theta} \right)\mathrm{d}\tau\right|^2
\end{align}
is the response of the detector to the zero mode in the static cylinder spacetime. Thus, in the simultaneous limit of late time and small universe size, Milne is again shown to recover the behaviour of the static cylinder.

Finally, it is worth noting that in several limiting regimes the zero mode of our periodic Milne resembles the infrared ambiguity present in the infinite Minkowski response. First, we recall the detector's response to this Minkowski ambiguity as calculated in appendix \ref{app:minkAsym},
\begin{align}
    \mathcal{F}^{m_0}(\omega) = (1-\ln(m_0\tau_f))\frac{\tau_f^2}{4\pi}\sinc^2 \! \left(\frac{\omega\tau_f}{2}\right) , 
\end{align}
where $m_0$ is the infrared cutoff. Note that this cutoff has a similar effect on the field theory, presenting an ambiguity that gives the system no well-defined Fock space. For the detector's response to the zero mode in periodic Milne \eqref{eq:zeroModeResponse}, taking either an early time, large universe limit, or simply taking the modulus of the free zero mode parameter $q$ to be large, we see a response
\begin{align}
    \mathcal{F}_{zm}\to\frac{p\tau_f^2}{32\sin\theta}\sinc^2 \! \left(\frac{\omega\tau_f}{2}\right) , 
\end{align}
which has the exact same dependence on system parameters as the infrared ambiguity in Minkowski, up to a relation between the zero mode parameter and the infrared cutoff they are identical.

\section{Numerical Results\label{sec:numerics}}

Via numerical evaluation, we show the response of the detector to have several notable features, as seen in figure~\ref{fig:asymComparison}. There is a smaller scale oscillation alongside a non-oscillatory component that gradually increases up until a peak near small negative $\mu$, before falling off quickly at positive $\mu$. This general shape could be expected since positive $\mu$ corresponds to excitation which requires energy and is therefore less likely than de-excitation, and is predicted by asymptotic expansion as evidenced by section \ref{subsec:largeSmallMu}. The response to the untwisted field also includes contributions from the zero mode, which provides a pronounced deviation from the twisted field.

We provide in appendix \ref{app:figures} several figures wherein a single system parameter is varied when compared to that of figure \ref{fig:asymComparison}, and in this way we gain insight into how these parameters impact the generic features of the detector's response. The impact of the zero mode ambiguity parameter $\zeta$, universe size parameter~$\xi$, interaction duration~$\sigma_f$, and detector peculiar velocity $v$ are all discussed in the remainder of this section.

\subsection{Zero Mode\label{subsec:zero mode numerics}}

We first turn our attention to the effect of the zero mode parameter on the detector's response to the untwisted field, fixing other parameters such that $\sigma_f = 1$, $\xi = 10$, $v=0$ identically to that already seen for the twisted case in figure \ref{fig:asymComparison}. The zero mode contribution included in figure \ref{fig:zeromode} is seen to have one peak at positive $\mu$ and one peak at negative $\mu$, approximately at the same $|\mu|$, with some smaller scale oscillations that interfere near $\mu=0$ but a rapid falloff such that the graph quickly approaches the same shape as the twisted response shown in figure \ref{fig:asymComparison}. This rapid falloff in the numerics is consistent with asymptotics calculated in appendices \ref{app:largePositiveMu}, \ref{app:largeNegativeMu}.

For large modulus of the zero mode parameter, $q$, the response is dominated by the $\sinc^2$ behaviour predicted in the analytical work of section \ref{subsec:spatialSizeLimits}. For small $q$, the symmetric two-peaked response dominates, similar to the massive case investigated in \cite{Toussaint2021}. The response of the detector to the zero mode is minimised for $q$ of order unity, but the exact value that minimises this response will likely depend on other system parameters.

For a fixed modulus $q$ of the zero mode parameter, changes in the phase $\theta$ determine which of the two peaks is more prominent. For $\theta$ close to zero the two peaks are of similar magnitude, but as $\theta$ increases to $\pi$ the peak at positive $\mu$ becomes comparatively larger. The small $\theta$ response diverges as could be expected from the analytical form of detector's response to the zero mode \eqref{eq:zeroModeResponse}, and the response is minimised for $\theta\approx 3\pi/2$.

\subsection{Universe Size}

The effect of varying the universe size parameter is seen in figure \ref{fig:xi}, which shows the response for selected values of~$\xi$, including the value $\xi=10$ of figure \ref{fig:asymComparison}. For negative $\mu$, the detector's response to the oscillatory modes show an insensitivity to the universe size parameter; even as the universe size triples the response is only shifted by approximately $15\%$. By contrast, positive $\mu$ exhibits a doubling of the peak height over this same parameter range. Thus, excitations are more sensitive to universe size changes than de-excitations. 

The response to the zero mode in the untwisted field shown in figure \ref{fig:xiuntwisted} is much more sensitive, when $\xi$ decreases towards $10$ the zero mode response starts to dominate even when the zero mode parameters are chosen to minimise it. This behaviour is as could be expected given the $1/\xi$ correspondence in the detector's response to the zero mode \eqref{eq:zeroModeResponse}, and the contrasting behaviour of the zero mode is in line with expectations calculated in section \ref{subsec:spatialSizeLimits}.

\subsection{Interaction Duration}

Figure \ref{fig:dsigma} displays the detector's response for a variety of durations of the interaction, increasing it from the value $\sigma_f=1$ of figure \ref{fig:asymComparison}.

For both the twisted and untwisted fields, as the duration increases the oscillation frequency of the response as a function of the gap parameter increases linearly, the only parameter that we have found to do so. This change can be expected from the asymptotic expansions calculated in appendix \ref{app:largePositiveMu}, since the next to leading order term in this expansion oscillates with a frequency linearly dependent on $\sigma_f$. The overall shape of the response also increases with the interaction duration, with the numerics showing that negative $\mu$ more closely approximates $\mu^{-1}$ for large $\sigma_f$, which is to be expected given the asymptotic expansion \eqref{eq:largeNegativeMu}.

Given that our switching function \eqref{eq:switching-choice} equals unity during the detector's operation, one expects the overall magnitude of the response to depend linearly on the duration of the interaction, and the numerical results for the twisted field in figure \ref{fig:dsigmatwisted} are consistent with this expectation. 
For the untwisted field, however, the contribution of the zero mode brings in additional effects: 
in figure \ref{fig:dsigmauntwisted}, over the parameter range probed therein,  
the detector's response approximately quintuples when the duration doubles. 

\subsection{Detector Velocity}

Peculiar velocities from $v=0$ to $v=0.9$ are investigated in figure \ref{fig:v}, and show the locations of the minima and maxima in the response of the detector to the oscillatory modes are only marginally affected by the change in observer velocity. By contrast, the maxima increase in height significantly, with the peaks closer to $\mu=0$ showing the most prominent changes. For sufficiently high velocities, the peak closest to $\mu=0$ becomes the maximum of the response, whereas the large $|\mu|$ peaks show less change. This could be expected since the large $|\mu|$ behaviour is identical to Minkowski, and it is therefore insensitive to changes in velocity.

The untwisted field's zero mode also provides interesting behaviour as shown in figure \ref{fig:vuntwisted}. The contribution from the zero mode is suppressed at higher velocities, showing that a fast observer sees a signal dominated by the contributions from the oscillatory field modes.

\section{Summary and discussion\label{sec:discussion}}

We have investigated a local quantum observer in an expanding $(1+1)$-dimensional Milne universe with compact spatial sections. 
We modelled the observer as a pointlike two-level system, an Unruh-DeWitt detector, 
coupled linearly to a real massless scalar field, 
with periodic or antiperiodic boundary conditions. 
The field was prepared in the conformal vacuum, 
except for the zero mode of the field with periodic boundary conditions, for which we therefore considered a family of Fock-type states labelled by a complex-valued parameter since it does not possess a conformal vacuum. 
We analysed how the transitions in an inertial Unruh-DeWitt detector 
depend on the age and size of the universe and the detector's peculiar velocity. 
In this scenario, the local quantum machinery enables a local detection of 
properties of the spacetime and the observer's motion 
that are classically accessible only via nonlocal observations. 

The detector's response as a function of the energy gap $\omega$ has the 
expected feature that excitations are suppressed compared with de-excitations,
and we showed analytically that the excitation probability has the large gap falloff $\mathcal{O}(\omega^{-2}\ln\omega)$, 
whereas the de-excitation probability has the weaker large gap falloff $\mathcal{O}(|\omega|^{-1})$. 
These falloffs agree with those for an inertial detector in Minkowski vacuum. 
For smaller $|\omega|$, the response exhibits significant smaller scale structure in the form of resonance peaks
whose locations and heights depend on the geometric parameters described above, as well as the duration of the interaction 
and, for the periodic field, on the choice of the zero mode parameter. 
We postpone the discussion of the zero mode contributions to the following paragraph. 
Without the zero mode contributions, the dependence of the response on the duration is broadly similar to that of an inertial detector in Minkowski vacuum, 
but varying the other parameters introduces other features. 
For example, our numerical plots show that the detector's peculiar velocity has only a minor effect on the locations of the resonance peaks, 
but the heights of the peaks increase significantly as the peculiar velocity increases, most notably at small~$|\omega|$. 
The numerical plots show subtle dependence on the age of the universe and on the spatial size, 
but one general feature, seen both numerically and analytically, is that the transition probabilities are suppressed when the spatial size decreases. 
In the limit of observations at late cosmological time, with spatial circumference at the observation time fixed, 
we recover the response of a detector on a static cylinder. 

Now return to the zero mode of the periodic field, which has no counterpart in Minkowski vacuum. 
We found that the zero mode contribution to the detector's response is very sensitive not only to the choice of the zero mode parameter, 
but also to the other system parameters. The zero mode contribution often dominates the total response, 
with a bimodal pair of excitation and de-excitation peaks at small values of~$\omega$, 
although the peaks become suppressed when the zero mode parameter is close to $1$ or the detector is moving fast.
The phase of the zero mode parameter affects which of the two peaks is more prominent, 
similarly to what happens with the corresponding mode for the massive field~\cite{Toussaint2021}.
In the small universe limit, the zero mode contribution diverges, completely dominating the oscillatory mode contributions which tend to zero. 
In the large universe limit, the zero mode contribution has a characteristic sinc squared divergence, 
similar to the infrared divergence in the response of an inertial detector in Minkowski vacuum. 
This sinc squared profile also emerges when the zero mode parameter is large and the other parameters are fixed. 

Our Unruh-DeWitt detector was switched on and off sharply. 
Whilst this allows the duration of the interaction to be defined cleanly, 
it has the side effect that the response bears some imprint 
of the kicks at the switch-on and switch-off moments. 
In four or more spacetime dimensions, the effect from the kicks would in fact be divergent \cite{Schlicht2004,Louko2007,Satz2007,Hodgkinson2012}.
A version of our analysis with a smoother switching would therefore be more readily comparable to that of a detector in four-dimensional spacetime. 
A version of our analysis for a four-dimensional Milne universe with compact spatial sections would however be qualitatively different in that the universe size parameters would be discrete rather than continuous,
because of the Mostow-Prasad rigidity theorem~\cite{Mostow1974}. 
Further, a four-dimensional Milne universe has limited applicability in the description of our own universe, 
because of its constant expansion rate \cite{Nielsen2016,Avelino2016,Tutusaus2017,Casado2020,Mohayaee2021}. 

In four-dimensional cosmological spacetimes that model our own universe more closely, 
an Unruh-DeWitt detector as a late time probe of early universe phenomena has been considered in \cite{Garay2014} 
in a setting with spatial flatness and $T^3$ spatial topology, with careful attention to the detector's switching, 
in a communications channel setting in~\cite{Blasco2016}, 
and in a non-Markovian memory setting in~\cite{Hsiang2021}: 
all of these analyses could be generalised to explore the effect of the five continuous shape parameters that the $T^3$ spatial topology admits \cite{Wolf2011,Louko:1989up}. 
All of these analyses could be further adapted to locally de Sitter cosmologies with $T^3$ spatial topology, as a model of the inflationary cosmological era, providing insights into the special issues that arise with a massless minimally-coupled field in de Sitter spacetime \cite{Allen1985,Allen1987,Kirsten1993}. 
Another direction of work would be to compare the detector's response with the field's renormalised stress-energy tensor, 
and specifically the contributions to each from the zero mode, as done for the massive field in $1+1$ dimensions in~\cite{Toussaint2021}. 
We leave these questions subject to future work.

\section*{Acknowledgments}

ASW thanks Nathaniel J. Roberts for assistance with the numerical results.
The work of JL was supported by United Kingdom Research and Innovation Science and Technology Facilities Council [grant numbers ST/S002227/1, ST/T006900/1 and ST/Y004523/1].
For the purpose of open access, the authors have applied a CC BY public copyright licence to any Author Accepted Manuscript version arising.

\clearpage
\appendix
\onecolumngrid
\section{Figures}\label{app:figures}

\begin{figure}[ht]
    \centering
    \subfloat[Selected values of the zero mode parameter magnitude $q$ ($\theta = \pi/4$)\label{fig:q}]{
        \includegraphics{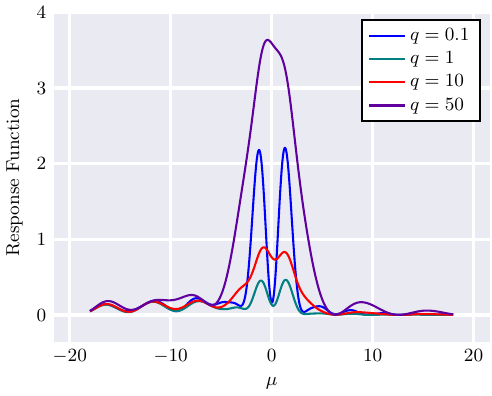}
    }
    \subfloat[Selected values of the zero mode parameter phase $\theta$ ($q = 1$)\label{fig:phi}]{
        \includegraphics{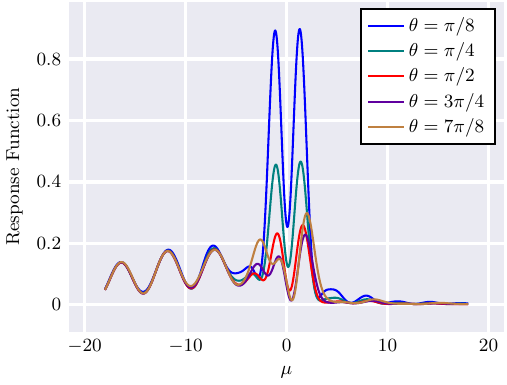}
    }
    \caption{
        The response function for the untwisted field as a function of the dimensionless energy $\mu$, with the parameter values $\xi = 10$, $\tau_f/t_0 = 1$ and $v=0$, and with selected values of the zero mode parameter $\zeta$ as shown. }
    \label{fig:zeromode}
\end{figure}

\begin{figure}[ht]
    \centering
    \subfloat[Detector response to the twisted field for selected universe size parameter values, $\xi$\label{fig:xitwisted}]{
        \includegraphics{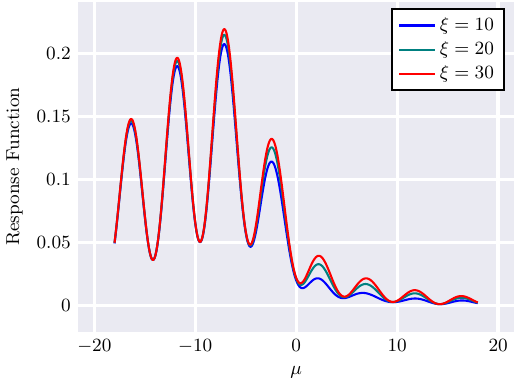}
    }
    \subfloat[Detector response to the untwisted field for selected universe size parameter values, $\xi$\label{fig:xiuntwisted}, with the zero mode parametrised by $q = 1$, $\theta = \pi/4$.]{
        \includegraphics{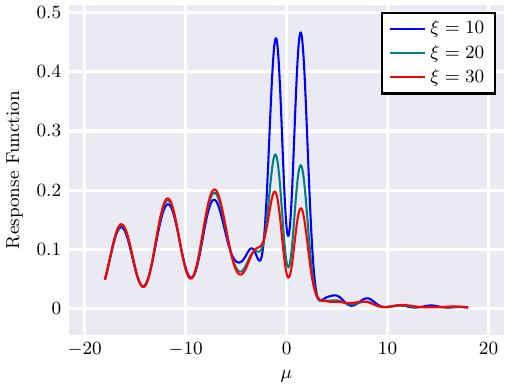}
    }
    \caption{The response function for the untwisted (a) and twisted (b) field as a function of the dimensionless energy $\mu$, with the parameter values $\tau_f/t_0 = 1$ and $v=0$, and with selected values of the size parameter $\xi$ as shown.}
    \label{fig:xi}
\end{figure}

\begin{figure}[ht]
    \centering
    \subfloat[Detector response to the twisted field for select values of detector interaction duration $\sigma_f$\label{fig:dsigmatwisted}]{
        \includegraphics{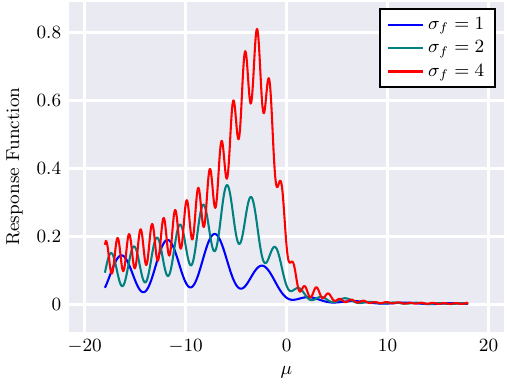}
    }
    \subfloat[Detector response to the untwisted field for select values of detector interaction duration $\sigma_f$, with the zero mode parametrised by $q = 1$, $\theta = \pi / 4$. \label{fig:dsigmauntwisted}]{
        \includegraphics{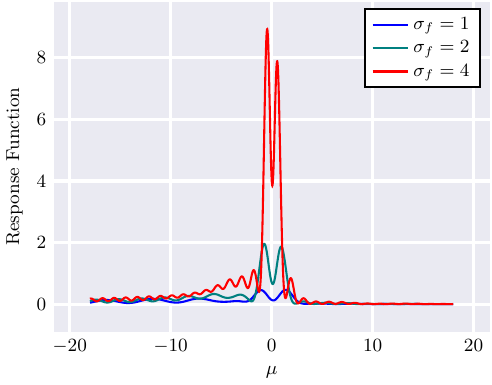}
    }
    \caption{The response function for the untwisted (a) and twisted (b) field as a function of the dimensionless energy $\mu$, with the parameter values $\xi = 10$, and $v=0$, and with selected values of the interaction duration $\sigma_f$ as shown.}
    \label{fig:dsigma}
\end{figure}

\begin{figure}[ht]
    \centering
    \subfloat[Detector response to the twisted field for select values of detector peculiar velocity $v$\label{fig:vtwisted}]{
        \includegraphics{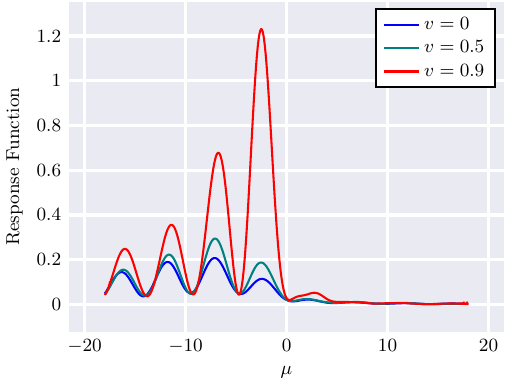}
    }
    \subfloat[Detector response to the untwisted field for select values of the detector peculiar velocity $v$, with the zero mode parametrised by $q = 1$, $\theta = \pi / 4$. \label{fig:vuntwisted}]{
        \includegraphics{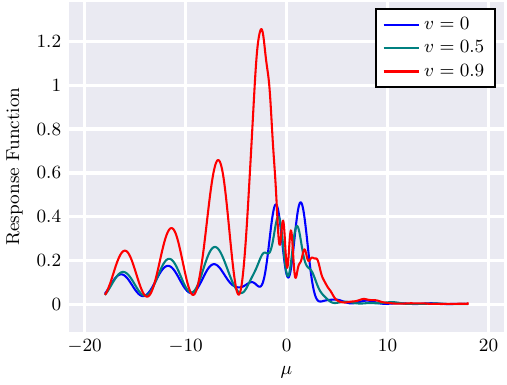}
    }
    \caption{The response function for the twisted (a) and untwisted (b) field as a function of the dimensionless energy $\mu$, with the parameter values $\xi = 10$, and $\sigma_f = 1$, and with selected values of the detector's peculiar velocity $v$ as shown.}
    \label{fig:v}
\end{figure}

\twocolumngrid
\clearpage

\section{Large Detector Excitation Gap}\label{app:largePositiveMu}

This section is dedicated to asymptotic expansion of the response of the detector \eqref{eq:detectorResponse} for large positive $\mu$. As such, we require a large $\mu$ asymptotic expansion of
\begin{subequations}
    \begin{align}
        \mathcal{F}(\mu) &= \frac{t_0^2}{2}\sum_{n=1}^\infty\frac{1}{M_n}\left( \left| f_n\left(\alpha_+\right) \right|^2 + \left| f_n\left(\alpha_-\right) \right|^2 \right)\label{eq:lpm-response} \\[3mm]
        f_n(\alpha, \mu) &= \int_\alpha^{\alpha+\sigma_f}\mathrm{d}x\exp\left[i\left(\mu x + \frac{M_n}{\xi}\ln(x)\right)\right]\label{eq:lpm-fn}\\[3mm]
        \alpha_\pm &= \sqrt\frac{1\pm v}{1\mp v} \\[3mm]
        M_n &= \begin{cases}
            2n\pi &     \mathrm{Untwisted}\\
            (2n-1)\pi & \mathrm{Twisted}.
        \end{cases}\label{eq:lpm-mn}
    \end{align}
\end{subequations}
First, define $r_n=M_n/\mu\xi$, $\beta=\alpha+\sigma_f$ and rewrite \eqref{eq:lpm-fn} with the substitution $y(x) = x + r_n\ln(x)$ to find the integral
\begin{align}
    f_n(\alpha, \mu) &= \int_{y(\alpha)}^{y(\beta)}\frac{x}{x+r_n}\mathrm{e}^{i\mu y}\mathrm{d}y
\end{align}
and repeatedly integrate by parts to find
\begin{subequations}
    \begin{align}
        f_{n}(\alpha, \mu) &= f_{n}^{(0)} + f_n^{(1)} + f_n^{(2)} \\
        f_{n}^{(0)} &= \left[\frac{x^{i\mu r_n + 1}\mathrm{e}^{i\mu x}}{i\mu (x + r_n)}\right]_\alpha^{\beta} \\
        f_{n}^{(1)} &= \frac{r_n}{\mu^2}\left[\frac{x^{i\mu r_n + 1}\mathrm{e}^{i\mu x}}{(x+r_n)^3}\right]_{\alpha}^{\beta} \\
        f_{n}^{(2)} &= -\frac{r_n}{\mu^2}\int_{y(\alpha)}^{y(\beta)}\frac{x(r_n-2x)}{(x+r_n)^5}\mathrm{e}^{i\mu y}\mathrm{d} y
    \end{align}
\end{subequations}

It can be shown that $|f_n^{(1)}| \leq c_1\mu^{-1}|f_n^{(0)}|$ and $|f_n^{(2)}| \leq c_2\mu^{-1}|f_n^{(0)}|$ where $c_1, c_2$ are independent of both $\mu$ and~$n$. It follows that for large positive $\mu$ we have
\begin{align}
|f_n| = |f_n^{(0)}|(1+\mathcal{O}(\mu^{-1})) ,
\label{eq:fn-uniformerror}
\end{align} 
where the error term is independent of~$n$.

Split \eqref{eq:lpm-response} into contributions from left and right movers such that
\begin{align}
    \mathcal{F}(\mu) &= S\left(\sqrt{\frac{1+v}{1-v}}, \mu\right) + S\left(\sqrt{\frac{1-v}{1+v}}, \mu\right)
\end{align}
and expand $S$ as
\begin{align}
    S(\alpha, \mu) &= \sum_{n=1}^\infty\frac{1}{M_n}\left| f_n(\alpha, \mu) \right|^2 \notag\\
        &= \sum_{n=1}^\infty\frac{1}{M_n}|f_n^{(0)}(\alpha, \mu)|^2\left( 1 + \mathcal{O}(\mu^{-1}) \right)^2 \notag\\
        &= (1+\mathcal{O}(\mu^{-1}))\sum_{n=1}^\infty\frac{1}{M_n}|f_n^{(0)}(\alpha, \mu)|^2 \notag\\
        &= (1+\mathcal{O}(\mu^{-1}))S^{(0)}(\alpha, \mu)
\end{align}
where the third equals sign follows from the $n$-independence of the error term in \eqref{eq:fn-uniformerror} and from the observation that $\left( 1 + \mathcal{O}(\mu^{-1}) \right)^2 = 1 + \mathcal{O}(\mu^{-1})$. 
This tells us that we can work with our leading order contribution $S^{(0)}$, but contributions from other terms will come into play at one order of $\mu$ lower. We can expand $S^{(0)}$ as
\begin{align}
      S^{(0)}&= \sum_{n=1}^\infty\frac{1}{M_n}\left| \left[\frac{x}{i\mu(x+r_n)}\mathrm{e}^{i\mu y(x)}\right]_\alpha^\beta\right|^2\notag\\
      &= \frac{1}{\mu^2}\sum_{n=1}^\infty\frac{1}{M_n}\biggl( \frac{\beta^2}{(\beta+r_n)^2} + \frac{\alpha^2}{(\alpha+r_n)^2} \notag\\&\hspace{4em}- \frac{2\alpha\beta\cos\left(\mu\sigma_f + M_n\lambda\right)}{(\beta+r_n)(\alpha+r_n)} \biggr) \notag\\
      &= \frac{1}{\mu^2}\bigl(S_1^{(0)}(\beta, \mu) + S_1^{(0)}(\alpha, \mu) \notag\\
        &\hspace{4em}+ S_2^{(0)}(\mu)\bigr)
\end{align}
where
\begin{align}
    \lambda := \frac{1}{\xi}\ln\left(\frac{\beta}{\alpha}\right).
\end{align}
To further characterise $S^{(0)}_{1,2}$, the twisted and untwisted fields must be evaluated separately from here onwards.

\subsection{Twisted Field}\label{lpm-twisted}

In pursuit of an expression of $S_1^{(0)}(x, \mu)$ in terms of known functions, define 
\begin{subequations}
    \begin{align}
        a &:= 2\pi/\xi\mu \\
        b &:= x + \pi/\xi\mu,
    \end{align}
\end{subequations}
offset the sum over $n$ by one such that our sum starts from zero, and split by partial fractions
\begin{align}
    S_1^{(0)}(x, \mu) &= \frac{x^2}{\pi}\sum_{n=0}^\infty\frac{1}{(2n+1)(an+b)^2} \notag\\
                &= \frac{x^2}{\pi}\frac{2}{(2b-a)^2}\Biggl[\sum_{n=0}^\infty\left(\frac{1}{n+1} - \frac{1}{n+b/a}\right)- \notag\\
                &\hspace{8em}\sum_{n=0}^\infty\left(\frac{1}{n+1} - \frac{1}{n + 1/2}\right)\Biggr] \notag\\
                &\hspace{3em}-\frac{x^2}{a\pi(2b-a)}\sum_{n=0}^\infty\frac{1}{(n+b/a)^2} \notag\\
                &= \frac{1}{2\pi}\left[\psi\left(\frac{b}{a}\right) - \psi\left(\frac{1}{2}\right) - \frac{2b-a}{2a}\psi'\left(\frac{b}{a}\right)\right] \notag\\[5mm]
                &= \frac{1}{2\pi}\biggl[ \psi\left(\frac{\mu\xi x+\pi}{2\pi}\right) - \frac{\mu\xi x}{2\pi}\psi'\left(\frac{\mu\xi x+\pi}{2\pi}\right) \notag\\&\hspace{2em} + 2\ln2 + \gamma \biggr],
\end{align}
where $\psi$ and $\psi'$ are the digamma \cite[Eq~5.2.2]{DLMF} and trigamma \cite[Eq~5.15.1]{DLMF} functions respectively, $\gamma$ is the Euler-Mascheroni constant arising from the known value of $\psi(1/2)$ \cite[Eq~5.4.13]{DLMF}, and the series expansion \cite[Eq~5.7.6]{DLMF} has been used to obtain the trigamma. For large $\mu$, make use of expansions \cite[Eq~2.4]{Ele} and \cite[Eq~5.15.8]{DLMF} to find
\begin{align}
    S_1^{(0)} (x, \mu)&= \frac{1}{2\pi}\left[\ln\mu + \ln\left(\frac{2\xi x}{\pi}\right) + \gamma - 1\right] + \mathcal{O}(\mu^{-2}).
\end{align}

To characterise $S_2^{(0)}$, define again $a := 2\pi/\xi\mu$, but newly define $c := \alpha + \pi/\xi\mu$, $d := \beta + \pi/\xi\mu$ and write
\begin{align}
    S_2^{(0)} (\mu) &= -2\alpha\beta\mathrm{Re}\left[\sum_{n=0}^\infty\frac{\mathrm{e}^{i\mu\sigma_f}\mathrm{e}^{i\pi\lambda}\left(\mathrm{e}^{2\pi i\lambda}\right)^n}{(2n+1)(an+c)(an+d)}\right].
\end{align}
We give $\lambda$ a small positive imaginary component $\pi\epsilon$ such that the sum converges, and will later take the $\epsilon\to0$ limit in the distributional sense. Splitting by partial fractions and using known functions, this can be rearranged to
\begin{align}
    S_2^{(0)} (\mu) &= -\mathrm{Re}\Biggl[ \frac{\mathrm{e}^{i\mu\sigma_f}\mathrm{e}^{i\pi\lambda - \epsilon}}{\pi\sigma_f}\biggl[ \alpha\mathcal{R}\left(\frac{\mu\xi\beta}{2\pi} + \frac{1}{2}, \lambda + i\frac{\epsilon}{\pi}, 1\right) \notag\\&\hspace{2em}- \beta\mathcal{R}\left(\frac{\mu\xi\alpha}{2\pi} + \frac{1}{2}, \lambda + i\epsilon/\pi, 1\right) \biggr] \Biggr] - \frac{s_2(\mu)}{\pi} , \label{lpm-twisted-s2}
\end{align}
where $\mathcal{R}(a, x, s)$ is the Lipschitz-Lerch zeta function as defined in \cite{Ferreira2004} and $s_2(\mu)$ is a more delicate term with complex branches. To find a precise form, write it as
\begin{align}
    s_2   &= \sum_{n=0}^\infty\frac{\mathrm{e}^{i\mu\sigma_f}\left(\mathrm{e}^{i\pi\lambda-\epsilon}\right)^{2n+1} + \mathrm{e}^{-i\mu\sigma_f}\left(\mathrm{e}^{-i\pi\lambda-\epsilon}\right)^{2n+1}}{2n+1} \notag\\
                &= \cos\left(\mu\sigma_f\right)\tanh^{-1}\left(\frac{\cos\left(\pi\lambda\right)}{\cosh\epsilon}\right) \notag\\
                &\hspace{2em}+ \frac{i}{2}\sin\left(\mu\sigma_f\right)\left[\ln\left(\frac{\sinh(\epsilon) + i\sin\left(\pi\lambda\right)}{\sinh(\epsilon) - i\sin(\pi\lambda)}\right)\right].
\end{align}
The first term is well-defined in the $\epsilon\to0$ limit, so we only need focus on the second. For $\sin(\pi\lambda) > 1$, the $\epsilon\to0$ limit gives $i\pi$, whereas for $\sin(\pi\lambda) < 1$ the limit yields $-i\pi$. It is therefore convenient to define
\begin{align}
    \Xi(x) &= \begin{cases}
        +1 & 2n < x < (2n+1) \\[3mm]
        -1 & (2n-1) < x < 2n 
    \end{cases} \hspace{2em} n \in \mathbb{N} \\
    &= \mathrm{sgn}\left(\sin\left(\pi x\right)\right)
\end{align}
such that
\begin{align}
    s_2(\mu) &= -\frac{1}{\pi}\cos\left(\mu\sigma_f\right)\ln\left| \tan\left(\frac{\pi\lambda}{2}\right) \right| - \frac{\pi\Xi(\lambda)}{2}\sin(\mu\sigma_f).
\end{align}
Note that this expansion is undefined when $\lambda$ is an integer. 

The Lipschitz-Lerch zeta functions in \eqref{lpm-twisted-s2} can be expanded using \cite[Eq.~7]{Ferreira2004}, which can be written as
\begin{align}
    \mathrm{e}^{i\pi\lambda}\mathcal{R}\left(x + \frac{1}{2}, \lambda, 1\right) = \frac{i}{2\sin\left(\pi\lambda\right)}\frac{1}{x} + \mathcal{O}(x^{-2})
\end{align}
and can be substituted back into \eqref{lpm-twisted-s2} to give
\begin{align}
    S_2^{(0)}(\mu)    &= \frac{1}{\pi}\cos\left(\mu\sigma_f\right)\ln\left|\tan\left(\frac{\pi\lambda}{2}\right)\right| + \frac{1}{2}\Xi(\lambda)\sin\left(\mu\sigma_f\right) \notag\\
                &\hspace{3em}-\frac{1}{\mu}\frac{2\beta}{\xi\alpha\beta}\frac{\sin(\mu\sigma_f)}{\sin(\pi\lambda)} - \mathcal{O}(\mu^{-2}).
\end{align}
This, along with the expansion of $S_1^{(0)}$, gives the final expansion
\begin{align}
    \frac{\mathcal{F}(\mu)}{t_0^2}  &= \frac{1}{\pi}\frac{\ln\mu}{\mu^2} \notag\\
                                    &+ \frac{1}{\mu^2}\frac{1}{2\pi}\Biggl[ \ln\left(\frac{4\xi^2}{\pi^2}\sqrt{\sigma_f^2 + 1 + \frac{2\sigma_f}{\sqrt{1-v^2}}}\right) + \gamma - 1 \notag\\
                                    &\hspace{1em}+\cos(\mu\sigma_f)\left[ \ln\left|\tan\left( \frac{\pi\lambda_+}{2} \right)\right| + \ln\left|\tan\left( \frac{\pi\lambda_-}{2} \right)\right|\right] \notag\\&\hspace{1em}+ \frac{\pi}{2}\sin(\mu\sigma_f)\left[ \Xi(\lambda_+) + \Xi(\lambda_-) \right]\Biggr] \notag\\
                                    &+ \mathcal{O}(\mu^{-3}),
\end{align}
where we expand the definition of $\lambda$ such that
\begin{align}
    \lambda_\pm &= \frac{1}{\xi}\ln\left(\frac{\alpha_\pm + \sigma_f}{\alpha_\pm}\right)\label{lpm-lambdapm}.
\end{align}

\subsection{Untwisted Field} \label{subapp:largePositiveMu-Untwisted}

For the untwisted field, $S_1^{(0)}$ takes the form
\begin{align}
    S_1^{(0)}(x, \mu) &= \frac{x^2}{2\pi}\sum_{n=1}^\infty\frac{1}{n(an+x)^2}
\end{align}
where $a = 2\pi/\xi\mu$. We again expand in terms of partial fractions
\begin{align}
    S_1^{(0)} &= \frac{1}{2\pi}\sum_{n=1}^\infty\left[\frac{1}{n} - \frac{1}{n+x/a}\right] \notag\\&\hspace{5em}- \frac{x}{2\pi a}\sum_{n=0}^\infty\frac{1}{(n+x/a)^2} + \frac{a}{2\pi x} \notag\\
                &= \frac{1}{2\pi}\left[ \psi\left( ax+1\right) + \gamma - ax\psi'\left(ax\right) \right] + \frac{a}{2\pi x} \notag\\
                &= \frac{1}{2\pi}\left[\ln\left(\frac{\mu\xi x}{2\pi}\right) + \gamma - 1\right] + \frac{1}{\xi\mu x} + \mathcal{O}(\mu^{-2}),
\end{align}
using the same expansions \cite[Eq~1.3]{Ele} and \cite[Eq~5.15.8]{DLMF} for trigamma and digamma functions.

The untwisted expression for $S_2^{(0)}(\mu)$ continues similarly to the twisted field, by the introduction of a small positive imaginary component to $\lambda$ we find
\begin{align}
    S_2^{(0)}(\mu)    &= -\frac{\mu^2\xi^2}{\pi}\mathrm{Re}\left[\sum_{n=1}^\infty\frac{\alpha\beta\mathrm{e}^{i\mu\sigma_f}\left(\mathrm{e}^{2\pi i\lambda - \epsilon}\right)^2}{n(2\pi n + \mu\xi\alpha)(2\pi n + \mu\xi\beta)}\right] \notag\\
                &= -\frac{1}{\pi}\mathrm{Re}\Biggl[\mathrm{e}^{i\mu\sigma_f}\biggl(\frac{\alpha}{\sigma_f}\sum_{n=1}^\infty\frac{\left(\mathrm{e}^{2\pi i\lambda - \epsilon}\right)^n}{n+w_1}\notag\\&\hspace{1em}+\frac{\beta}{\sigma_f}\sum_{n=1}^\infty\frac{\left(\mathrm{e}^{2\pi i\lambda-\epsilon}\right)^n}{n+w_2}+ \frac{\alpha}{\sigma_f}\sum_{n=1}^\infty\frac{\left(\mathrm{e}^{2\pi i\lambda-\epsilon}\right)^n}{n}\biggr)\Biggr] \notag\\
                &= -\frac{1}{\pi}\mathrm{Re}\Biggl[\mathrm{e}^{i\mu\sigma_f}\biggl(\frac{\alpha}{\sigma_f}\mathcal{R}\left(w_1, \lambda + i\epsilon/2\pi, 1\right)\notag\\&\hspace{4em}- \frac{\beta}{\sigma_f}\mathcal{R}\left(w_2, \lambda + i\epsilon/2\pi, 1\right) \notag\\&\hspace{4em}+ \frac{2\pi}{\mu\xi}\left(\frac{1}{\alpha} + \frac{1}{\beta}\right)\biggr)\Biggr] + \frac{1}{\pi}s_2(\mu)
\end{align}
where the elementary term comes from inserting the $n=0$ sum terms, and the new variables are defined by
\begin{align}
    w_1 &= \frac{\mu\xi\beta}{2\pi} & w_2 &= \frac{\mu\xi\alpha}{2\pi}.
\end{align}

Again, $s_2$ is a more delicate term with complex branching behaviour,
\begin{align}
    s_2   &= -\mathrm{Re}\left[\mathrm{e}^{i\mu\sigma_f}\sum_{n=1}^\infty\frac{\left(\mathrm{e}^{2\pi i\lambda - \epsilon}\right)^n}{n}\right] \notag\\
                &= \mathrm{Re}\left[\mathrm{e}^{i\mu\sigma_f}\ln\left(1-\mathrm{e}^{2\pi i\lambda}\mathrm{e}^{-\epsilon}\right)\right] \notag\\
                &= \mathrm{Re}\left[\mathrm{e}^{i\mu\sigma_f}\ln\left(1-b\cos\theta-ib\sin\theta\right)\right]
\end{align}
where two constants have been introduced for simplicity, $b$ is a positive constant that will be $1$ in the limiting case and $\theta=2\pi\lambda$. To maintain the branches of the logarithm, we calculate the argument of the logarithm for $b = 1$ and ensure it is in the same quadrant of the complex plane as for $b < 1$. For $b=1$, this argument is
\begin{align}
    x = 2\left|\sin\frac{\theta}{2}\right|\exp\left(\frac{i\theta}{2} + i\left(n - \frac{1}{2}\right)\pi\right)
\end{align}
for some integer $n$. This must be in the 4th quadrant for $\sin\theta > 0$, and enforcing this restraint gives $n = -\lfloor\lambda\rfloor$, where $\lfloor x\rfloor$ is the floor function. This must also be in the 1st quadrant for $\sin\theta < 0$ and on the real line for $\sin\theta = 0$, both of which give the same restraint for the full range of $\theta$, and therefore $\lambda$, values. The expansion is undefined when $\lambda$ takes integer values, similarly to the twisted case.

Now it is possible to take the principal valued logarithm as $\epsilon\to 0$ and calculate
\begin{align}
    s_2(\mu)  &= \mathrm{Re}\left[ \mathrm{e}^{i\mu\sigma_f}\ln\left(2|\sin(\pi\lambda)|\mathrm{e}^{i\pi\left(\{\lambda\} - \frac{1}{2}\right)}\right) \right] \notag\\
                    &= \cos(\mu\sigma_f)\ln(2|\sin(\pi\lambda)|) \notag\\&\hspace{4em}+ \pi\sin(\mu\sigma_f)\left(\{\lambda\} - \frac{1}{2}\right)
\end{align}
where $\{x\}$ denotes the fractional part of $x$. The full expression of $S_2$ is therefore
\begin{align}
S_2    &= -\frac{1}{\pi}\mathrm{Re}\Biggl[\mathrm{e}^{i\mu\sigma_f}\Biggl(\frac{\alpha}{\sigma_f}\mathcal{R}\left(w_1, \lambda, 1\right)- \frac{\beta}{\sigma_f}\mathcal{R}\left(w_2, \lambda, 1\right) \notag\\&\hspace{3em}+ \frac{2\pi}{\mu\xi}\left(\frac{1}{\alpha} + \frac{1}{\beta}\right)\Biggr)\Biggr]\!+\!\frac{1}{\pi}\cos(\mu\sigma_f)\ln(2|\sin(n\lambda)|) \notag\\&\hspace{3em}+ \sin(\mu\sigma_f)\left(\{\lambda\} - \frac{1}{2}\right).
\end{align}

The untwisted field also has the notable addition of the zero mode \eqref{eq:zeroModeResponse}, at which point we note that the exponential integrals are $\mathcal{O}(\mu^{-1})$ \cite{Wong2001}, and therefore do not contribute to order $\mathcal{O}(\mu^{-2})$ in the response. The zero mode will, however, have contributions at $\mathcal{O}(\mu^{-2})$ originating from the other terms in \eqref{eq:zeroModeResponse}.

Thus, the untwisted field has similar but distinct large positive $\mu$ behaviour of
\begin{align}
    \frac{\mathcal{F}(\mu)}{t_0^2} &= \frac{1}{\pi}\frac{\ln\mu}{\mu^2} \notag\\
                                    &+\frac{1}{\mu^2\pi}\Biggl[\ln\left(\frac{4\xi^2}{\pi^2}\sqrt{\sigma_f^2 + 1 + \frac{2\sigma_f}{\sqrt{1-v^2}}}\right) + \gamma - 1 \notag\\
                                    &\hspace{1em}+\cos(\mu\sigma_f)\ln\left[2|\sin(\pi\lambda_-)||\sin(\pi\lambda_+)|\right] \notag\\&\hspace{1em}+ \sin(\mu\sigma_f)\left(\{\lambda_-\} + \{\lambda_+\} - 1\right)\Biggr] \notag\\
                                    &+\frac{\left| \ln\bigl[ (\sigma_f + \alpha^+)(\sigma_f + \alpha^-) \bigr] - 2q\mathrm{e}^{i\theta}(\mathrm{e}^{i\mu\sigma_f} - 1) \right|^2}{8\xi q\sin\theta\mu^2} \notag\\
                                    &+\mathcal{O}(\mu^{-3})
\end{align}
where~$\lambda_\pm$ is defined similarly by \eqref{lpm-lambdapm}, and the last written term is the contribution from the zero mode.

\section{Large De-Excitation Gap}\label{app:largeNegativeMu}

Large de-excitation gaps correspond to large negative~$\omega$. Our starting point is identical to appendix \ref{app:largePositiveMu}, with our oscillatory response function
\begin{subequations}
    \begin{align}
        \mathcal{F}(\mu) &= \frac{t_0^2}{2}\left[\sum_{n=1}^\infty\frac{1}{M_n}\left( \left| f_n^+\left(\mu\right) \right|^2 + \left| f_n^-\left(\mu\right) \right|^2 \right)\label{eq:lnm-response}\right] \\[3mm]
        f_n^\pm(\mu) &= \int_{\alpha_\pm}^{\beta_\pm}\mathrm{d}x\exp\left[i\left(\mu x + \frac{M_n}{\xi}\ln(x)\right)\right]\label{eq:lnm-fn}\\[3mm]
        \alpha_\pm &= \sqrt{\frac{1\pm v}{1\mp v}} \hspace{5em} \beta_\pm = \alpha_\pm + \sigma_f\\[3mm]
        M_n &= \begin{cases}
            2n\pi &     \mathrm{Untwisted}\\
            (2n-1)\pi & \mathrm{Twisted},
        \end{cases}\label{eq:lnm-mn}
    \end{align}
\end{subequations}
however a difference will present itself when it comes to the substitution
\begin{align}
    y = x + \frac{M_n}{\xi\mu}\ln(x).
\end{align}
To illustrate this, define $\Lambda = -\mu > 0$ and write
\begin{align}
    y = x - \frac{M_n}{\xi\Lambda}\ln(x) = x - r_n\ln(x)
\end{align}
which is a two-to-one relation, and so the substitution needs to be treated more carefully. Note three distinct domains covering \(r_n\) a.e.,
\begin{subequations}
    \begin{align}
        {\rm (I)}& \quad r_n < \alpha_\pm \\
        {\rm (II)}& \quad \alpha_\pm < r_n < \beta_\pm \\
        {\rm (III)}& \quad r_n > \beta_\pm,
    \end{align}
\end{subequations}
where we discard $r_n = \alpha_\pm$, $\beta_\pm$, finding that \(f^\pm\) has distinct behaviour in each region. Again, split the response function \eqref{eq:lnm-response} into left and right moving field modes
\begin{align}
    \mathcal{F}(\mu) &= \frac{t_0^2}{2}\left[S\left(\sqrt\frac{1+v}{1-v}, \mu\right) + S\left(\sqrt{\frac{1-v}{1+v}}, \mu\right)\right]
\end{align}
defining
\begin{align}
    S(\alpha, \mu) &= \sum_{n=1}^\infty\frac{\left| f_n(\alpha) \right|^2}{M_n} \notag\\
    &= \sum_{n=1}^{r_n < \alpha}\frac{\left| f_n(\alpha) \right|^2}{M_n} + \sum_{r_n > \alpha}^{r_n < \beta}\frac{\left| f_n(\alpha) \right|^2}{M_n} + \sum_{r_n > \beta}^\infty\frac{\left| f_n(\alpha) \right|^2}{M_n} \notag\\
    &= S_{\rm (I)} + S_{\rm (II)} + S_{\rm (III)}.
\end{align}
$S_{(\rm I)}$ and $S_{(\rm III)}$ continue in a very similar fashion to appendix \ref{app:largePositiveMu}, giving expansions of digamma and trigamma functions, 
and 
giving a final form of both $S_1$ and $S_3$ $\mathcal{O}(\mu^{-2} \ln|\mu|)$. The untwisted zero mode is $\mathcal{O}(\mu^{-2})$ as was calculated in appendix \ref{subapp:largePositiveMu-Untwisted}. The finite sum $S_{\rm (II)}$ involves the integral
\begin{align}
    I(\mu) &= \int_\alpha^\beta\mathrm{d}x \exp\left[-i\Lambda\left(x - r_n\ln(x)\right)\right]
\end{align}
and since $\alpha < r_n < \beta$ for $S_2$, this integral contains a single minimum in the phase at $x = r_n$. Thus, methods of stationary phase such as \cite[Eq.~3.2]{Wong2001} can be used to find
\begin{align}
    I(\mu) &= \sqrt\frac{2\pi r_n}{\Lambda}\exp\left[i\left(\Lambda r_n\left(1 - \ln r_n\right) - \frac{\pi}{4}\right)\right] + \mathcal{O}(\Lambda^{-1})
\end{align}
giving
\begin{align}
    S_{\rm (II)} = \sum_{r_n > \alpha}^{r_n < \beta}\frac{2\pi r_n}{\Lambda M_n} + \mathcal{O}(\Lambda^{-2}) = \frac{2\pi}{\Lambda^2\xi}\sum_{r_n > \alpha}^{r_n < \beta}1 + \mathcal{O}(\Lambda^{-2}).
\end{align}
The number of terms in this sum is
\begin{align}
    N = \frac{\xi\Lambda(\beta - \alpha)}{2\pi} = \frac{\xi\Lambda\sigma_f}{2\pi},
\end{align}
and so we get the final expression for $S_{\rm (II)}$
\begin{align}
    S_{\rm (II)} = \frac{\sigma_f}{\Lambda} + \mathcal{O}(\mu^{-2}).
\end{align}
Since this is the dominant term in the expansion, the full expansion for $\mathcal{F}$ at large negative $\mu$ can be found as
\begin{align}
    \mathcal{F}(\mu) &= \frac{t_0^2\sigma_f}{|\mu|} + \mathcal{O}(\mu^{-2}\ln|\mu|).
\end{align}

\section{Large Excitation and De-Excitation Gaps in Minkowski\label{app:minkAsym}} 

In this appendix, we look at the large positive and negative gap asymptotics of the response of a detector in (1+1) Minkowski, given by
\begin{subequations}
\label{eq:minkResp}
\begin{align}
    \mathcal{F}(\omega) &= \int_0^\infty\mathrm{d}s\left[\sin(\omega s) + \frac{2}{\pi}\cos(\omega s)\ln(m_0 s)\right]I_\chi(s), 
    \label{eq:minkResp-Fonly} \\
    I_\chi(s) &= -\frac{1}{2}\int_{-\infty}^\infty\mathrm{d}u\chi(u)\chi(u - s), 
    \label{eq:minkResp-Ifunction}
\end{align}
\end{subequations}
as obtained in \cite[Eq~3.3a]{Louko2014}, where $m_0$ parametrises the infrared ambiguity present for non-derivative coupling. 

As the Minkowski vacuum is invariant under translations, the interaction does not have a distinguished starting time, 
and we may therefore take the switching function to be 
\begin{align}
    \chi(\tau) = \Theta\left(\frac{\Delta}{2} + \tau\right)\Theta\left(\frac{\Delta}{2} - \tau\right) , 
\end{align}
where $\Delta$ is the duration of the interaction. 
From \eqref{eq:minkResp-Ifunction} we then have 
\begin{align}
    I_\chi(s) = -\frac{1}{2}\Theta(\Delta - s)\Theta(\Delta + s)(\Delta - |s|).
\label{eq:Ichi-triangle}
\end{align}
Substituting \eqref{eq:Ichi-triangle} in~\eqref{eq:minkResp-Fonly}, we obtain 
\begin{subequations}
\begin{align}
\mathcal{F}(\omega) &= \mathcal{F}^{(s)}(\omega) + \mathcal{F}^{(c)}(\omega) , \\
\mathcal{F}^{(s)}(\omega) & = -\frac{1}{2}\int_0^{\Delta}\mathrm{d}s\sin\left(\omega s\right)(\Delta - s), 
\label{eq:Mink-Fs-int}\\
\mathcal{F}^{(c)}(\omega) &= - \frac{1}{\pi}\int_0^{\Delta} \cos\left(\omega s\right)\ln (m_0 s)\;(\Delta - s) . 
\label{eq:Mink-Fc-int}
\end{align}
\end{subequations}

$\mathcal{F}^{(s)}$ \eqref{eq:Mink-Fs-int} evaluates to
\begin{align}
    \mathcal{F}^{(s)}(\omega) &= - \frac{\Delta}{2\omega} + \frac{1}{2\omega^2}\sin(\omega\Delta). 
\end{align}
To evaluate $\mathcal{F}^{(c)}$ \eqref{eq:Mink-Fc-int}, we first integrate by parts, integrating the trigonometric factor, with the result
\begin{align}
    \mathcal{F}^{(c)}(\omega) &= \frac{1}{\omega\pi}\int_0^{\Delta}\mathrm{d}s\sin(\omega s)\left(\frac{\Delta - s}{s} - \ln (m_0 s)\right) \notag\\
                      &= \frac{\Delta}{\omega\pi}\Si\left(\omega\Delta\right) 
                      + \frac{\Delta^2}{2\pi}\sinc^2 \! \left(\frac{\omega\Delta}{2}\right) \notag\\
                      &\hspace{2ex}-\frac{1}{\omega\pi}\int_{0}^{\Delta}\mathrm{d}s\sin(\omega s)\ln (m_0s) , 
\label{eq:Mink-Fc-intt}
\end{align}
where in the last equality we have used the definition of the sine integral \cite[Eq.~6.2.9]{DLMF}. 
In the remaining integral in \eqref{eq:Mink-Fc-intt}, we again integrate by parts, using 
$\sin(\omega s) = \frac{d}{ds} \! \left[\frac{1}{\omega}\bigl(1 - \cos(\omega s)\bigr) \right]$, 
obtaining 
\begin{align}
    \mathcal{F}^{(c)}(\omega) &= \frac{\Delta}{\omega\pi}\Si\left(\omega\Delta\right) + \frac{1}{\omega^2\pi}\Cin(\omega\Delta) \notag\\
    &\hspace{3ex}+ \frac{\Delta^2}{2\pi}\sinc^2 \! \left(\frac{\omega\Delta}{2}\right) \bigl(1-\ln(m_0\Delta)\bigr) , 
\end{align}
using the definition of the cosine integral \cite[Eq.~6.2.12]{DLMF}. 

Collecting, and using \cite[Eq.~6.2.10]{DLMF}, we have
\begin{align}
    \mathcal{F}(\omega) &= \frac{\Delta}{\pi\omega}\si(\omega\Delta) + \frac{1}{\pi\omega^2}\Cin(\omega\Delta) + \frac{\sin(\omega\Delta)}{2\omega^2} \notag\\
                                &\hspace{3ex}+ \frac{\Delta^2}{2\pi}\sinc^2 \! \left(\frac{\omega\Delta}{2}\right)\bigl(1-\ln(m_0\Delta)\bigr).
\end{align}
By the large argument asymptotics of $\si$ and $\Cin$ \cite{DLMF}, 
$\mathcal{F}(\omega)$ has the large positive $\omega$ asymptotics
\begin{align}
    \mathcal{F}(\omega) &= \frac{\ln(\omega\Delta)}{\pi\omega^2} + \mathcal{O}(\omega^{-2})
\end{align}
and the large negative $\omega$ asymptotics
\begin{align}
    \mathcal{F}(\omega) &= \frac{\Delta}{|\omega|} 
    + \frac{\ln(|\omega|\Delta)}{\pi\omega^2}
    + \mathcal{O}(\omega^{-2}).
\end{align}

\bibliography{References}

\begin{thebibliography}{54}%
\makeatletter
\providecommand \@ifxundefined [1]{%
 \@ifx{#1\undefined}
}%
\providecommand \@ifnum [1]{%
 \ifnum #1\expandafter \@firstoftwo
 \else \expandafter \@secondoftwo
 \fi
}%
\providecommand \@ifx [1]{%
 \ifx #1\expandafter \@firstoftwo
 \else \expandafter \@secondoftwo
 \fi
}%
\providecommand \natexlab [1]{#1}%
\providecommand \enquote  [1]{``#1''}%
\providecommand \bibnamefont  [1]{#1}%
\providecommand \bibfnamefont [1]{#1}%
\providecommand \citenamefont [1]{#1}%
\providecommand \href@noop [0]{\@secondoftwo}%
\providecommand \href [0]{\begingroup \@sanitize@url \@href}%
\providecommand \@href[1]{\@@startlink{#1}\@@href}%
\providecommand \@@href[1]{\endgroup#1\@@endlink}%
\providecommand \@sanitize@url [0]{\catcode `\\12\catcode `\$12\catcode
  `\&12\catcode `\#12\catcode `\^12\catcode `\_12\catcode `\%12\relax}%
\providecommand \@@startlink[1]{}%
\providecommand \@@endlink[0]{}%
\providecommand \url  [0]{\begingroup\@sanitize@url \@url }%
\providecommand \@url [1]{\endgroup\@href {#1}{\urlprefix }}%
\providecommand \urlprefix  [0]{URL }%
\providecommand \Eprint [0]{\href }%
\providecommand \doibase [0]{https://doi.org/}%
\providecommand \selectlanguage [0]{\@gobble}%
\providecommand \bibinfo  [0]{\@secondoftwo}%
\providecommand \bibfield  [0]{\@secondoftwo}%
\providecommand \translation [1]{[#1]}%
\providecommand \BibitemOpen [0]{}%
\providecommand \bibitemStop [0]{}%
\providecommand \bibitemNoStop [0]{.\EOS\space}%
\providecommand \EOS [0]{\spacefactor3000\relax}%
\providecommand \BibitemShut  [1]{\csname bibitem#1\endcsname}%
\let\auto@bib@innerbib\@empty
\bibitem [{\citenamefont {Reeh}\ and\ \citenamefont
  {Schlieder}(1961)}]{Reeh1961}%
  \BibitemOpen
  \bibfield  {author} {\bibinfo {author} {\bibfnamefont {H.}~\bibnamefont
  {Reeh}}\ and\ \bibinfo {author} {\bibfnamefont {S.}~\bibnamefont
  {Schlieder}},\ }\bibfield  {title} {\bibinfo {title} {{Bemerkungen zur
  {Unit\"ar\"aquivalenz} von lorentzinvarianten {Feldern}}},\ }\href
  {https://doi.org/10.1007/BF02787889} {\bibfield  {journal} {\bibinfo
  {journal} {Nuovo Cimento}\ }\textbf {\bibinfo {volume} {22}},\ \bibinfo
  {pages} {1051} (\bibinfo {year} {1961})}\BibitemShut {NoStop}%
\bibitem [{\citenamefont {Birrell}\ and\ \citenamefont
  {Davies}(1982)}]{Birrell1982}%
  \BibitemOpen
  \bibfield  {author} {\bibinfo {author} {\bibfnamefont {N.~D.}\ \bibnamefont
  {Birrell}}\ and\ \bibinfo {author} {\bibfnamefont {P.~C.~W.}\ \bibnamefont
  {Davies}},\ }\href {https://books.google.co.uk/books?id=YYxzQgAACAAJ} {\emph
  {\bibinfo {title} {Quantum Fields in Curved Space}}},\ Cambridge Monographs
  on Mathematical Physics\ (\bibinfo  {publisher} {Cambridge University
  Press},\ \bibinfo {year} {1982})\BibitemShut {NoStop}%
\bibitem [{\citenamefont {Mukhanov}\ and\ \citenamefont
  {Winitzki}(2007)}]{Mukhanov2007}%
  \BibitemOpen
  \bibfield  {author} {\bibinfo {author} {\bibfnamefont {V.}~\bibnamefont
  {Mukhanov}}\ and\ \bibinfo {author} {\bibfnamefont {S.}~\bibnamefont
  {Winitzki}},\ }\href@noop {} {\emph {\bibinfo {title} {Introduction to
  Quantum Effects in Gravity}}}\ (\bibinfo  {publisher} {Cambridge University
  Press},\ \bibinfo {year} {2007})\BibitemShut {NoStop}%
\bibitem [{\citenamefont {Levin.}(2002)}]{Levin2001}%
  \BibitemOpen
  \bibfield  {author} {\bibinfo {author} {\bibfnamefont {J.~J.}\ \bibnamefont
  {Levin.}},\ }\bibfield  {title} {\bibinfo {title} {{Topology and the cosmic
  microwave background}},\ }\href
  {https://doi.org/10.1016/S0370-1573(02)00018-2} {\bibfield  {journal}
  {\bibinfo  {journal} {Physics Reports}\ }\textbf {\bibinfo {volume} {365}},\
  \bibinfo {pages} {251} (\bibinfo {year} {2002})}\BibitemShut {NoStop}%
\bibitem [{\citenamefont {Ford}\ and\ \citenamefont
  {Pathinayake}(1989)}]{Ford1989}%
  \BibitemOpen
  \bibfield  {author} {\bibinfo {author} {\bibfnamefont {L.~H.}\ \bibnamefont
  {Ford}}\ and\ \bibinfo {author} {\bibfnamefont {C.}~\bibnamefont
  {Pathinayake}},\ }\bibfield  {title} {\bibinfo {title} {Bosonic
  zero-frequency modes and initial conditions.},\ }\href
  {https://doi.org/10.1103/PhysRevD.39.3642} {\bibfield  {journal} {\bibinfo
  {journal} {Physical Review D}\ }\textbf {\bibinfo {volume} {39}},\ \bibinfo
  {pages} {3642} (\bibinfo {year} {1989})}\BibitemShut {NoStop}%
\bibitem [{\citenamefont {Toussaint}\ and\ \citenamefont
  {Louko}(2021)}]{Toussaint2021}%
  \BibitemOpen
  \bibfield  {author} {\bibinfo {author} {\bibfnamefont {V.}~\bibnamefont
  {Toussaint}}\ and\ \bibinfo {author} {\bibfnamefont {J.}~\bibnamefont
  {Louko}},\ }\bibfield  {title} {\bibinfo {title} {Detecting the massive
  bosonic zero-mode in expanding cosmological spacetimes},\ }\href
  {https://doi.org/10.1103/physrevd.103.105011} {\bibfield  {journal} {\bibinfo
   {journal} {Physical Review D}\ }\textbf {\bibinfo {volume} {103}},\ \bibinfo
  {pages} {105011} (\bibinfo {year} {2021})}\BibitemShut {NoStop}%
\bibitem [{\citenamefont {Garay}\ \emph {et~al.}(2014)\citenamefont {Garay},
  \citenamefont {Martín-Benito},\ and\ \citenamefont
  {Martín-Martínez}}]{Garay2014}%
  \BibitemOpen
  \bibfield  {author} {\bibinfo {author} {\bibfnamefont {L.~J.}\ \bibnamefont
  {Garay}}, \bibinfo {author} {\bibfnamefont {M.}~\bibnamefont
  {Martín-Benito}},\ and\ \bibinfo {author} {\bibfnamefont {E.}~\bibnamefont
  {Martín-Martínez}},\ }\bibfield  {title} {\bibinfo {title} {Echo of the
  quantum bounce},\ }\href {https://doi.org/10.1103/physrevd.89.043510}
  {\bibfield  {journal} {\bibinfo  {journal} {Physical Review D}\ }\textbf
  {\bibinfo {volume} {89}},\ \bibinfo {pages} {043510} (\bibinfo {year}
  {2014})}\BibitemShut {NoStop}%
\bibitem [{\citenamefont {Toussaint}\ and\ \citenamefont
  {Louko}(2024)}]{Toussaint2024}%
  \BibitemOpen
  \bibfield  {author} {\bibinfo {author} {\bibfnamefont {V.}~\bibnamefont
  {Toussaint}}\ and\ \bibinfo {author} {\bibfnamefont {J.}~\bibnamefont
  {Louko}},\ }\bibfield  {title} {\bibinfo {title} {Vacua in locally de sitter
  cosmologies, and how to distinguish them},\ }\href
  {https://doi.org/10.1103/physrevd.109.025007} {\bibfield  {journal} {\bibinfo
   {journal} {Physical Review D}\ }\textbf {\bibinfo {volume} {109}},\ \bibinfo
  {pages} {025007} (\bibinfo {year} {2024})}\BibitemShut {NoStop}%
\bibitem [{\citenamefont {Milne}(1932)}]{Milne1932nature}%
  \BibitemOpen
  \bibfield  {author} {\bibinfo {author} {\bibfnamefont {E.~A.}\ \bibnamefont
  {Milne}},\ }\bibfield  {title} {\bibinfo {title} {World structure and the
  expansion of the universe},\ }\href@noop {} {\bibfield  {journal} {\bibinfo
  {journal} {Nature}\ }\textbf {\bibinfo {volume} {130}},\ \bibinfo {pages} {9}
  (\bibinfo {year} {1932})}\BibitemShut {NoStop}%
\bibitem [{\citenamefont {Rindler}(2006)}]{Rindler:2006km}%
  \BibitemOpen
  \bibfield  {author} {\bibinfo {author} {\bibfnamefont {W.}~\bibnamefont
  {Rindler}},\ }\href@noop {} {\emph {\bibinfo {title} {{Relativity: Special,
  general, and cosmological}}}}\ (\bibinfo  {publisher} {Oxford University
  Press},\ \bibinfo {year} {2006})\BibitemShut {NoStop}%
\bibitem [{\citenamefont {Nielsen}\ \emph {et~al.}(2016)\citenamefont
  {Nielsen}, \citenamefont {Guffanti},\ and\ \citenamefont
  {Sarkar}}]{Nielsen2016}%
  \BibitemOpen
  \bibfield  {author} {\bibinfo {author} {\bibfnamefont {J.~T.}\ \bibnamefont
  {Nielsen}}, \bibinfo {author} {\bibfnamefont {A.}~\bibnamefont {Guffanti}},\
  and\ \bibinfo {author} {\bibfnamefont {S.}~\bibnamefont {Sarkar}},\
  }\bibfield  {title} {\bibinfo {title} {Marginal evidence for cosmic
  acceleration from type {Ia} supernovae},\ }\href
  {https://doi.org/10.1038/srep35596} {\bibfield  {journal} {\bibinfo
  {journal} {Scientific Reports}\ }\textbf {\bibinfo {volume} {6}},\ \bibinfo
  {pages} {35596} (\bibinfo {year} {2016})}\BibitemShut {NoStop}%
\bibitem [{\citenamefont {Avelino}\ and\ \citenamefont
  {Kirshner}(2016)}]{Avelino2016}%
  \BibitemOpen
  \bibfield  {author} {\bibinfo {author} {\bibfnamefont {A.}~\bibnamefont
  {Avelino}}\ and\ \bibinfo {author} {\bibfnamefont {R.~P.}\ \bibnamefont
  {Kirshner}},\ }\bibfield  {title} {\bibinfo {title} {The dimensionless age of
  the universe: A riddle for our time},\ }\href
  {https://doi.org/10.3847/0004-637X/828/1/35} {\bibfield  {journal} {\bibinfo
  {journal} {The Astrophysical Journal}\ }\textbf {\bibinfo {volume} {828}},\
  \bibinfo {pages} {35} (\bibinfo {year} {2016})}\BibitemShut {NoStop}%
\bibitem [{\citenamefont {Tutusaus}\ \emph {et~al.}(2017)\citenamefont
  {Tutusaus}, \citenamefont {Lamine}, \citenamefont {Dupays},\ and\
  \citenamefont {Blanchard}}]{Tutusaus2017}%
  \BibitemOpen
  \bibfield  {author} {\bibinfo {author} {\bibfnamefont {I.}~\bibnamefont
  {Tutusaus}}, \bibinfo {author} {\bibfnamefont {B.}~\bibnamefont {Lamine}},
  \bibinfo {author} {\bibfnamefont {A.}~\bibnamefont {Dupays}},\ and\ \bibinfo
  {author} {\bibfnamefont {A.}~\bibnamefont {Blanchard}},\ }\bibfield  {title}
  {\bibinfo {title} {Is cosmic acceleration proven by local cosmological
  probes?},\ }\href {https://doi.org/10.1051/0004-6361/201630289} {\bibfield
  {journal} {\bibinfo  {journal} {Astronomy and Astrophysics}\ }\textbf
  {\bibinfo {volume} {602}},\ \bibinfo {pages} {A73} (\bibinfo {year}
  {2017})}\BibitemShut {NoStop}%
\bibitem [{\citenamefont {Casado}(2020)}]{Casado2020}%
  \BibitemOpen
  \bibfield  {author} {\bibinfo {author} {\bibfnamefont {J.}~\bibnamefont
  {Casado}},\ }\bibfield  {title} {\bibinfo {title} {Linear expansion models
  vs. standard cosmologies: a critical and historical overview},\ }\href
  {https://doi.org/10.1007/s10509-019-3720-z} {\bibfield  {journal} {\bibinfo
  {journal} {Astrophysics and Space Science}\ }\textbf {\bibinfo {volume}
  {365}},\ \bibinfo {eid} {16} (\bibinfo {year} {2020})}\BibitemShut {NoStop}%
\bibitem [{\citenamefont {Mohayaee}\ \emph {et~al.}(2021)\citenamefont
  {Mohayaee}, \citenamefont {Rameez},\ and\ \citenamefont
  {Sarkar}}]{Mohayaee2021}%
  \BibitemOpen
  \bibfield  {author} {\bibinfo {author} {\bibfnamefont {R.}~\bibnamefont
  {Mohayaee}}, \bibinfo {author} {\bibfnamefont {M.}~\bibnamefont {Rameez}},\
  and\ \bibinfo {author} {\bibfnamefont {S.}~\bibnamefont {Sarkar}},\
  }\bibfield  {title} {\bibinfo {title} {Do supernovae indicate an accelerating
  universe?},\ }\href {https://doi.org/10.1140/epjs/s11734-021-00199-6}
  {\bibfield  {journal} {\bibinfo  {journal} {The European Physical Journal
  Special Topics}\ }\textbf {\bibinfo {volume} {230}},\ \bibinfo {pages}
  {2067–2076} (\bibinfo {year} {2021})}\BibitemShut {NoStop}%
\bibitem [{\citenamefont {Unruh}(1976)}]{Unruh1976}%
  \BibitemOpen
  \bibfield  {author} {\bibinfo {author} {\bibfnamefont {W.~G.}\ \bibnamefont
  {Unruh}},\ }\bibfield  {title} {\bibinfo {title} {Notes on black-hole
  evaporation},\ }\href {https://doi.org/10.1103/PhysRevD.14.870} {\bibfield
  {journal} {\bibinfo  {journal} {Physical Review D}\ }\textbf {\bibinfo
  {volume} {14}},\ \bibinfo {pages} {870} (\bibinfo {year} {1976})}\BibitemShut
  {NoStop}%
\bibitem [{\citenamefont {DeWitt}(1979)}]{DeWitt1979}%
  \BibitemOpen
  \bibfield  {author} {\bibinfo {author} {\bibfnamefont {B.~S.}\ \bibnamefont
  {DeWitt}},\ }\bibfield  {title} {\bibinfo {title} {Quantum gravity: the new
  synthesis},\ }in\ \href@noop {} {\emph {\bibinfo {booktitle} {General
  Relativity: an Einstein centenary survey}}},\ \bibinfo {editor} {edited by\
  \bibinfo {editor} {\bibfnamefont {S.~W.}\ \bibnamefont {Hawking}}\ and\
  \bibinfo {editor} {\bibfnamefont {W.}~\bibnamefont {Israel}}}\ (\bibinfo
  {publisher} {Cambridge University Press},\ \bibinfo {address} {Cambridge},\
  \bibinfo {year} {1979})\BibitemShut {NoStop}%
\bibitem [{\citenamefont {Martín-Martínez}\ \emph {et~al.}(2013)\citenamefont
  {Martín-Martínez}, \citenamefont {Montero},\ and\ \citenamefont {del
  Rey}}]{Martinez2013}%
  \BibitemOpen
  \bibfield  {author} {\bibinfo {author} {\bibfnamefont {E.}~\bibnamefont
  {Martín-Martínez}}, \bibinfo {author} {\bibfnamefont {M.}~\bibnamefont
  {Montero}},\ and\ \bibinfo {author} {\bibfnamefont {M.}~\bibnamefont {del
  Rey}},\ }\bibfield  {title} {\bibinfo {title} {Wavepacket detection with the
  {Unruh}-{DeWitt} model},\ }\href {https://doi.org/10.1103/physrevd.87.064038}
  {\bibfield  {journal} {\bibinfo  {journal} {Physical Review D}\ }\textbf
  {\bibinfo {volume} {87}},\ \bibinfo {pages} {064038} (\bibinfo {year}
  {2013})}\BibitemShut {NoStop}%
\bibitem [{\citenamefont {Alhambra}\ \emph {et~al.}(2014)\citenamefont
  {Alhambra}, \citenamefont {Kempf},\ and\ \citenamefont
  {Mart\'{\i}n-Mart\'{\i}nez}}]{Alhambra2014}%
  \BibitemOpen
  \bibfield  {author} {\bibinfo {author} {\bibfnamefont {{\'A}.~M.}\
  \bibnamefont {Alhambra}}, \bibinfo {author} {\bibfnamefont {A.}~\bibnamefont
  {Kempf}},\ and\ \bibinfo {author} {\bibfnamefont {E.}~\bibnamefont
  {Mart\'{\i}n-Mart\'{\i}nez}},\ }\bibfield  {title} {\bibinfo {title} {Casimir
  forces on atoms in optical cavities},\ }\href
  {https://doi.org/10.1103/PhysRevA.89.033835} {\bibfield  {journal} {\bibinfo
  {journal} {Physical Review A}\ }\textbf {\bibinfo {volume} {89}},\ \bibinfo
  {pages} {033835} (\bibinfo {year} {2014})}\BibitemShut {NoStop}%
\bibitem [{\citenamefont {Rajaraman}(1982)}]{Rajaraman1982}%
  \BibitemOpen
  \bibfield  {author} {\bibinfo {author} {\bibfnamefont {R.}~\bibnamefont
  {Rajaraman}},\ }\href@noop {} {\emph {\bibinfo {title} {Solitons and
  Instantons. An Introduction to Solitons and Instantons in Quantum Field
  Theory}}}\ (\bibinfo  {publisher} {North-Holland Publishing Company},\
  \bibinfo {year} {1982})\BibitemShut {NoStop}%
\bibitem [{\citenamefont {Allen}\ and\ \citenamefont
  {Folacci}(1987)}]{Allen1987}%
  \BibitemOpen
  \bibfield  {author} {\bibinfo {author} {\bibfnamefont {B.}~\bibnamefont
  {Allen}}\ and\ \bibinfo {author} {\bibfnamefont {A.}~\bibnamefont
  {Folacci}},\ }\bibfield  {title} {\bibinfo {title} {Massless minimally
  coupled scalar field in de {Sitter} space},\ }\href
  {https://doi.org/10.1103/PhysRevD.35.3771} {\bibfield  {journal} {\bibinfo
  {journal} {Physical Review D}\ }\textbf {\bibinfo {volume} {35}},\ \bibinfo
  {pages} {3771} (\bibinfo {year} {1987})}\BibitemShut {NoStop}%
\bibitem [{\citenamefont {Garriga}\ and\ \citenamefont
  {Vilenkin}(1992)}]{Garriga1992}%
  \BibitemOpen
  \bibfield  {author} {\bibinfo {author} {\bibfnamefont {J.}~\bibnamefont
  {Garriga}}\ and\ \bibinfo {author} {\bibfnamefont {A.}~\bibnamefont
  {Vilenkin}},\ }\bibfield  {title} {\bibinfo {title} {Quantum fluctuations on
  domain walls, strings, and vacuum bubbles},\ }\href
  {https://doi.org/10.1103/PhysRevD.45.3469} {\bibfield  {journal} {\bibinfo
  {journal} {Physical Review D}\ }\textbf {\bibinfo {volume} {45}},\ \bibinfo
  {pages} {3469} (\bibinfo {year} {1992})}\BibitemShut {NoStop}%
\bibitem [{\citenamefont {McCartor}\ and\ \citenamefont
  {Robertson}(1992)}]{McCartor1992}%
  \BibitemOpen
  \bibfield  {author} {\bibinfo {author} {\bibfnamefont {G.}~\bibnamefont
  {McCartor}}\ and\ \bibinfo {author} {\bibfnamefont {D.~G.}\ \bibnamefont
  {Robertson}},\ }\bibfield  {title} {\bibinfo {title} {Bosonic zero modes in
  discretized light cone field theory},\ }\href
  {https://doi.org/10.1007/BF01559747} {\bibfield  {journal} {\bibinfo
  {journal} {Zeitschrift f\"ur Physik C}\ }\textbf {\bibinfo {volume} {53}},\
  \bibinfo {pages} {679} (\bibinfo {year} {1992})}\BibitemShut {NoStop}%
\bibitem [{\citenamefont {Kirsten}\ and\ \citenamefont
  {Garriga}(1993)}]{Kirsten1993}%
  \BibitemOpen
  \bibfield  {author} {\bibinfo {author} {\bibfnamefont {K.}~\bibnamefont
  {Kirsten}}\ and\ \bibinfo {author} {\bibfnamefont {J.}~\bibnamefont
  {Garriga}},\ }\bibfield  {title} {\bibinfo {title} {Massless minimally
  coupled fields in de {Sitter} space: O(4)-symmetric states versus de
  {Sitter}--invariant vacuum},\ }\href
  {https://doi.org/10.1103/PhysRevD.48.567} {\bibfield  {journal} {\bibinfo
  {journal} {Physical Review D}\ }\textbf {\bibinfo {volume} {48}},\ \bibinfo
  {pages} {567} (\bibinfo {year} {1993})}\BibitemShut {NoStop}%
\bibitem [{\citenamefont {Martín-Martínez}\ and\ \citenamefont
  {Louko}(2014)}]{Martinez2014}%
  \BibitemOpen
  \bibfield  {author} {\bibinfo {author} {\bibfnamefont {E.}~\bibnamefont
  {Martín-Martínez}}\ and\ \bibinfo {author} {\bibfnamefont {J.}~\bibnamefont
  {Louko}},\ }\bibfield  {title} {\bibinfo {title} {Particle detectors and the
  zero mode of a quantum field},\ }\href
  {https://doi.org/10.1103/physrevd.90.024015} {\bibfield  {journal} {\bibinfo
  {journal} {Physical Review D}\ }\textbf {\bibinfo {volume} {90}},\ \bibinfo
  {pages} {024015} (\bibinfo {year} {2014})}\BibitemShut {NoStop}%
\bibitem [{\citenamefont {Louko}\ and\ \citenamefont
  {Toussaint}(2016)}]{Louko2016}%
  \BibitemOpen
  \bibfield  {author} {\bibinfo {author} {\bibfnamefont {J.}~\bibnamefont
  {Louko}}\ and\ \bibinfo {author} {\bibfnamefont {V.}~\bibnamefont
  {Toussaint}},\ }\bibfield  {title} {\bibinfo {title} {{Unruh}-{DeWitt}
  detector's response to fermions in flat spacetimes},\ }\href
  {https://doi.org/10.1103/physrevd.94.064027} {\bibfield  {journal} {\bibinfo
  {journal} {Physical Review D}\ }\textbf {\bibinfo {volume} {94}},\ \bibinfo
  {pages} {064027} (\bibinfo {year} {2016})}\BibitemShut {NoStop}%
\bibitem [{\citenamefont {Tjoa}\ and\ \citenamefont
  {Martín-Martínez}(2020)}]{Tjoa2020}%
  \BibitemOpen
  \bibfield  {author} {\bibinfo {author} {\bibfnamefont {E.}~\bibnamefont
  {Tjoa}}\ and\ \bibinfo {author} {\bibfnamefont {E.}~\bibnamefont
  {Martín-Martínez}},\ }\bibfield  {title} {\bibinfo {title} {Vacuum
  entanglement harvesting with a zero mode},\ }\href
  {https://doi.org/10.1103/physrevd.101.125020} {\bibfield  {journal} {\bibinfo
   {journal} {Physical Review D}\ }\textbf {\bibinfo {volume} {101}},\ \bibinfo
  {pages} {125020} (\bibinfo {year} {2020})}\BibitemShut {NoStop}%
\bibitem [{\citenamefont {Milne}(1936)}]{Milne1936}%
  \BibitemOpen
  \bibfield  {author} {\bibinfo {author} {\bibfnamefont {E.~A.}\ \bibnamefont
  {Milne}},\ }\bibfield  {title} {\bibinfo {title} {Relativity, gravitation,
  and world-structure},\ }\href@noop {} {\bibfield  {journal} {\bibinfo
  {journal} {Philosophy}\ }\textbf {\bibinfo {volume} {11}} (\bibinfo {year}
  {1936})}\BibitemShut {NoStop}%
\bibitem [{\citenamefont {Walker}(1937)}]{Walker1937}%
  \BibitemOpen
  \bibfield  {author} {\bibinfo {author} {\bibfnamefont {A.~G.}\ \bibnamefont
  {Walker}},\ }\bibfield  {title} {\bibinfo {title} {On {Milne}'s theory of
  world-structure},\ }\href@noop {} {\bibfield  {journal} {\bibinfo  {journal}
  {Proceedings of the London Mathematical Society}\ }\textbf {\bibinfo {volume}
  {2}},\ \bibinfo {pages} {90} (\bibinfo {year} {1937})}\BibitemShut {NoStop}%
\bibitem [{\citenamefont {Rubin}\ and\ \citenamefont
  {Hayden}(2016)}]{Rubin2016}%
  \BibitemOpen
  \bibfield  {author} {\bibinfo {author} {\bibfnamefont {D.}~\bibnamefont
  {Rubin}}\ and\ \bibinfo {author} {\bibfnamefont {B.}~\bibnamefont {Hayden}},\
  }\bibfield  {title} {\bibinfo {title} {Is the expansion of the universe
  expanding? {All} signs point to yes},\ }\href
  {https://doi.org/10.3847/2041-8213/833/2/l30} {\bibfield  {journal} {\bibinfo
   {journal} {The Astrophysical Journal Letters}\ }\textbf {\bibinfo {volume}
  {833}},\ \bibinfo {pages} {L30} (\bibinfo {year} {2016})}\BibitemShut
  {NoStop}%
\bibitem [{\citenamefont {Batra}\ \emph {et~al.}(2000)\citenamefont {Batra},
  \citenamefont {Lohiya}, \citenamefont {Mahajan},\ and\ \citenamefont
  {Muhkerjee}}]{Batra2000}%
  \BibitemOpen
  \bibfield  {author} {\bibinfo {author} {\bibfnamefont {A.}~\bibnamefont
  {Batra}}, \bibinfo {author} {\bibfnamefont {D.}~\bibnamefont {Lohiya}},
  \bibinfo {author} {\bibfnamefont {S.}~\bibnamefont {Mahajan}},\ and\ \bibinfo
  {author} {\bibfnamefont {A.}~\bibnamefont {Muhkerjee}},\ }\bibfield  {title}
  {\bibinfo {title} {Nucleosynthesis in a universe with a linearly evolving
  scale factor},\ }\href {https://doi.org/10.1142/S0218271800000682} {\bibfield
   {journal} {\bibinfo  {journal} {International Journal of Modern Physics D}\
  }\textbf {\bibinfo {volume} {09}},\ \bibinfo {pages} {757} (\bibinfo {year}
  {2000})}\BibitemShut {NoStop}%
\bibitem [{\citenamefont {Gehlaut}\ \emph {et~al.}(2003)\citenamefont
  {Gehlaut}, \citenamefont {Kumar}, \citenamefont {Geetanjali},\ and\
  \citenamefont {Lohiya}}]{Gehlaut2003}%
  \BibitemOpen
  \bibfield  {author} {\bibinfo {author} {\bibfnamefont {S.}~\bibnamefont
  {Gehlaut}}, \bibinfo {author} {\bibfnamefont {P.}~\bibnamefont {Kumar}},
  \bibinfo {author} {\bibnamefont {Geetanjali}},\ and\ \bibinfo {author}
  {\bibfnamefont {D.}~\bibnamefont {Lohiya}},\ }\href@noop {} {\bibinfo {title}
  {A concordant ``freely coasting cosmology"}} (\bibinfo {year} {2003}),\
  \Eprint {https://arxiv.org/abs/astro-ph/0306448} {arXiv:astro-ph/0306448
  [astro-ph]} \BibitemShut {NoStop}%
\bibitem [{\citenamefont {Hawking}\ and\ \citenamefont
  {Ellis}(1973)}]{Hawking1973}%
  \BibitemOpen
  \bibfield  {author} {\bibinfo {author} {\bibfnamefont {S.~W.}\ \bibnamefont
  {Hawking}}\ and\ \bibinfo {author} {\bibfnamefont {G.~F.~R.}\ \bibnamefont
  {Ellis}},\ }\href@noop {} {\emph {\bibinfo {title} {The Large Scale Structure
  of Space-Time}}},\ Cambridge Monographs on Mathematical Physics\ (\bibinfo
  {publisher} {Cambridge University Press},\ \bibinfo {year}
  {1973})\BibitemShut {NoStop}%
\bibitem [{\citenamefont {Uzan}\ \emph {et~al.}(2002)\citenamefont {Uzan},
  \citenamefont {Luminet}, \citenamefont {Lehoucq},\ and\ \citenamefont
  {Peter}}]{Uzan2002}%
  \BibitemOpen
  \bibfield  {author} {\bibinfo {author} {\bibfnamefont {J.-P.}\ \bibnamefont
  {Uzan}}, \bibinfo {author} {\bibfnamefont {J.-P.}\ \bibnamefont {Luminet}},
  \bibinfo {author} {\bibfnamefont {R.}~\bibnamefont {Lehoucq}},\ and\ \bibinfo
  {author} {\bibfnamefont {P.}~\bibnamefont {Peter}},\ }\bibfield  {title}
  {\bibinfo {title} {The twin paradox and space topology},\ }\href
  {https://doi.org/10.1088/0143-0807/23/3/306} {\bibfield  {journal} {\bibinfo
  {journal} {European Journal of Physics}\ }\textbf {\bibinfo {volume} {23}},\
  \bibinfo {pages} {277} (\bibinfo {year} {2002})}\BibitemShut {NoStop}%
\bibitem [{\citenamefont {Dray}(1990)}]{Dray1990}%
  \BibitemOpen
  \bibfield  {author} {\bibinfo {author} {\bibfnamefont {T.}~\bibnamefont
  {Dray}},\ }\bibfield  {title} {\bibinfo {title} {The twin paradox
  revisited},\ }\href@noop {} {\bibfield  {journal} {\bibinfo  {journal}
  {American Journal of Physics}\ }\textbf {\bibinfo {volume} {58}},\ \bibinfo
  {pages} {822} (\bibinfo {year} {1990})}\BibitemShut {NoStop}%
\bibitem [{\citenamefont {Low}(1990)}]{Low1990}%
  \BibitemOpen
  \bibfield  {author} {\bibinfo {author} {\bibfnamefont {R.~J.}\ \bibnamefont
  {Low}},\ }\bibfield  {title} {\bibinfo {title} {An acceleration-free version
  of the clock paradox},\ }\href {https://doi.org/10.1088/0143-0807/11/1/003}
  {\bibfield  {journal} {\bibinfo  {journal} {European Journal of Physics}\
  }\textbf {\bibinfo {volume} {11}},\ \bibinfo {pages} {25} (\bibinfo {year}
  {1990})}\BibitemShut {NoStop}%
\bibitem [{\citenamefont {Weinberg}\ \emph {et~al.}(2013)\citenamefont
  {Weinberg}, \citenamefont {Mortonson}, \citenamefont {Eisenstein},
  \citenamefont {Hirata}, \citenamefont {Riess},\ and\ \citenamefont
  {Rozo}}]{Weinberg2013}%
  \BibitemOpen
  \bibfield  {author} {\bibinfo {author} {\bibfnamefont {D.~H.}\ \bibnamefont
  {Weinberg}}, \bibinfo {author} {\bibfnamefont {M.~J.}\ \bibnamefont
  {Mortonson}}, \bibinfo {author} {\bibfnamefont {D.~J.}\ \bibnamefont
  {Eisenstein}}, \bibinfo {author} {\bibfnamefont {C.}~\bibnamefont {Hirata}},
  \bibinfo {author} {\bibfnamefont {A.~G.}\ \bibnamefont {Riess}},\ and\
  \bibinfo {author} {\bibfnamefont {E.}~\bibnamefont {Rozo}},\ }\bibfield
  {title} {\bibinfo {title} {{Observational Probes of Cosmic Acceleration}},\
  }\href {https://doi.org/10.1016/j.physrep.2013.05.001} {\bibfield  {journal}
  {\bibinfo  {journal} {Physics Reports}\ }\textbf {\bibinfo {volume} {530}},\
  \bibinfo {pages} {87} (\bibinfo {year} {2013})}\BibitemShut {NoStop}%
\bibitem [{\citenamefont {Higuchi}\ \emph {et~al.}(2023)\citenamefont
  {Higuchi}, \citenamefont {Schmieding},\ and\ \citenamefont
  {Blanco}}]{Higuchi:2022nfy}%
  \BibitemOpen
  \bibfield  {author} {\bibinfo {author} {\bibfnamefont {A.}~\bibnamefont
  {Higuchi}}, \bibinfo {author} {\bibfnamefont {L.}~\bibnamefont
  {Schmieding}},\ and\ \bibinfo {author} {\bibfnamefont {D.~S.}\ \bibnamefont
  {Blanco}},\ }\bibfield  {title} {\bibinfo {title} {{Automorphic scalar fields
  in two-dimensional de Sitter space}},\ }\href
  {https://doi.org/10.1088/1361-6382/aca73f} {\bibfield  {journal} {\bibinfo
  {journal} {Classical and Quantum Gravity}\ }\textbf {\bibinfo {volume}
  {40}},\ \bibinfo {pages} {015009} (\bibinfo {year} {2023})}\BibitemShut
  {NoStop}%
\bibitem [{\citenamefont {Junker}\ and\ \citenamefont
  {Schrohe}(2002)}]{Junker:2001gx}%
  \BibitemOpen
  \bibfield  {author} {\bibinfo {author} {\bibfnamefont {W.}~\bibnamefont
  {Junker}}\ and\ \bibinfo {author} {\bibfnamefont {E.}~\bibnamefont
  {Schrohe}},\ }\bibfield  {title} {\bibinfo {title} {{Adiabatic vacuum states
  on general space-time manifolds: Definition, construction, and physical
  properties}},\ }\href {https://doi.org/10.1007/s000230200001} {\bibfield
  {journal} {\bibinfo  {journal} {Annales Henri Poincar\'e}\ }\textbf {\bibinfo
  {volume} {3}},\ \bibinfo {pages} {1113} (\bibinfo {year} {2002})}\BibitemShut
  {NoStop}%
\bibitem [{\citenamefont {Schlicht}(2004)}]{Schlicht2004}%
  \BibitemOpen
  \bibfield  {author} {\bibinfo {author} {\bibfnamefont {S.}~\bibnamefont
  {Schlicht}},\ }\bibfield  {title} {\bibinfo {title} {Considerations on the
  {Unruh} effect: causality and regularization},\ }\href
  {https://doi.org/10.1088/0264-9381/21/19/011} {\bibfield  {journal} {\bibinfo
   {journal} {Classical and Quantum Gravity}\ }\textbf {\bibinfo {volume}
  {21}},\ \bibinfo {pages} {4647} (\bibinfo {year} {2004})}\BibitemShut
  {NoStop}%
\bibitem [{\citenamefont {Louko}\ and\ \citenamefont {Satz}(2007)}]{Louko2007}%
  \BibitemOpen
  \bibfield  {author} {\bibinfo {author} {\bibfnamefont {J.}~\bibnamefont
  {Louko}}\ and\ \bibinfo {author} {\bibfnamefont {A.}~\bibnamefont {Satz}},\
  }\bibfield  {title} {\bibinfo {title} {Excited by a quantum field: does shape
  matter?},\ }\href {https://doi.org/10.1088/1742-6596/68/1/012014} {\bibfield
  {journal} {\bibinfo  {journal} {Journal of Physics: Conference Series}\
  }\textbf {\bibinfo {volume} {68}},\ \bibinfo {pages} {012014} (\bibinfo
  {year} {2007})}\BibitemShut {NoStop}%
\bibitem [{\citenamefont {Satz}(2007)}]{Satz2007}%
  \BibitemOpen
  \bibfield  {author} {\bibinfo {author} {\bibfnamefont {A.}~\bibnamefont
  {Satz}},\ }\bibfield  {title} {\bibinfo {title} {Then again, how often does
  the {Unruh}–{DeWitt} detector click if we switch it carefully?},\ }\href
  {https://doi.org/10.1088/0264-9381/24/7/003} {\bibfield  {journal} {\bibinfo
  {journal} {Classical and Quantum Gravity}\ }\textbf {\bibinfo {volume}
  {24}},\ \bibinfo {pages} {1719} (\bibinfo {year} {2007})}\BibitemShut
  {NoStop}%
\bibitem [{\citenamefont {Hodgkinson}\ and\ \citenamefont
  {Louko}(2012)}]{Hodgkinson2012}%
  \BibitemOpen
  \bibfield  {author} {\bibinfo {author} {\bibfnamefont {L.}~\bibnamefont
  {Hodgkinson}}\ and\ \bibinfo {author} {\bibfnamefont {J.}~\bibnamefont
  {Louko}},\ }\bibfield  {title} {\bibinfo {title} {How often does the
  {Unruh}-{DeWitt} detector click beyond four dimensions?},\ }\href
  {https://doi.org/10.1063/1.4739453} {\bibfield  {journal} {\bibinfo
  {journal} {Journal of Mathematical Physics}\ }\textbf {\bibinfo {volume}
  {53}},\ \bibinfo {pages} {082301} (\bibinfo {year} {2012})}\BibitemShut
  {NoStop}%
\bibitem [{DLM(2024)}]{DLMF}%
  \BibitemOpen
  \href {https://dlmf.nist.gov/} {\bibinfo {title} {\it {NIST} digital library
  of mathematical functions}},\ \bibinfo {howpublished}
  {\url{https://dlmf.nist.gov/}, Release 1.2.0 of 2024-03-15} (\bibinfo {year}
  {2024}),\ \bibinfo {note} {{F}.~W.~J. Olver, A.~B. {Olde Daalhuis}, D.~W.
  Lozier, B.~I. Schneider, R.~F. Boisvert, C.~W. Clark, B.~R. Miller, B.~V.
  Saunders, H.~S. Cohl, and M.~A. McClain, eds.}\BibitemShut {Stop}%
\bibitem [{\citenamefont {Louko}(2014)}]{Louko2014}%
  \BibitemOpen
  \bibfield  {author} {\bibinfo {author} {\bibfnamefont {J.}~\bibnamefont
  {Louko}},\ }\bibfield  {title} {\bibinfo {title} {{Unruh}-{DeWitt} detector
  response across a {Rindler} firewall is finite},\ }\href
  {https://doi.org/10.1007/jhep09(2014)142} {\bibfield  {journal} {\bibinfo
  {journal} {Journal of High Energy Physics}\ }\textbf {\bibinfo {volume}
  {2014}},\ \bibinfo {pages} {142} (\bibinfo {year} {2014})}\BibitemShut
  {NoStop}%
\bibitem [{\citenamefont {Mostow}(1974)}]{Mostow1974}%
  \BibitemOpen
  \bibfield  {author} {\bibinfo {author} {\bibfnamefont {G.~D.}\ \bibnamefont
  {Mostow}},\ }\href {https://doi.org/doi:10.1515/9781400881833} {\emph
  {\bibinfo {title} {Strong Rigidity of Locally Symmetric Spaces}}},\ \bibinfo
  {series} {Annals of Mathematical Studies}, Vol.~\bibinfo {volume} {78}\
  (\bibinfo  {publisher} {Princeton University Press},\ \bibinfo {address}
  {Princeton},\ \bibinfo {year} {1974})\BibitemShut {NoStop}%
\bibitem [{\citenamefont {Blasco}\ \emph {et~al.}(2016)\citenamefont {Blasco},
  \citenamefont {Garay}, \citenamefont {Mart\'{\i}n-Benito},\ and\
  \citenamefont {Mart\'{\i}n-Mart\'{\i}nez}}]{Blasco2016}%
  \BibitemOpen
  \bibfield  {author} {\bibinfo {author} {\bibfnamefont {A.}~\bibnamefont
  {Blasco}}, \bibinfo {author} {\bibfnamefont {L.~J.}\ \bibnamefont {Garay}},
  \bibinfo {author} {\bibfnamefont {M.}~\bibnamefont {Mart\'{\i}n-Benito}},\
  and\ \bibinfo {author} {\bibfnamefont {E.}~\bibnamefont
  {Mart\'{\i}n-Mart\'{\i}nez}},\ }\bibfield  {title} {\bibinfo {title}
  {Timelike information broadcasting in cosmology},\ }\href
  {https://doi.org/10.1103/PhysRevD.93.024055} {\bibfield  {journal} {\bibinfo
  {journal} {Physical Review D}\ }\textbf {\bibinfo {volume} {93}},\ \bibinfo
  {pages} {024055} (\bibinfo {year} {2016})}\BibitemShut {NoStop}%
\bibitem [{\citenamefont {Hsiang}\ and\ \citenamefont {Hu}(2021)}]{Hsiang2021}%
  \BibitemOpen
  \bibfield  {author} {\bibinfo {author} {\bibfnamefont {J.-T.}\ \bibnamefont
  {Hsiang}}\ and\ \bibinfo {author} {\bibfnamefont {B.-L.}\ \bibnamefont
  {Hu}},\ }\bibfield  {title} {\bibinfo {title} {{NonMarkovianity in cosmology:
  Memories kept in a quantum field}},\ }\href
  {https://doi.org/10.1016/j.aop.2021.168656} {\bibfield  {journal} {\bibinfo
  {journal} {Annals Physics}\ }\textbf {\bibinfo {volume} {434}},\ \bibinfo
  {pages} {168656} (\bibinfo {year} {2021})}\BibitemShut {NoStop}%
\bibitem [{\citenamefont {Wolf}(2011)}]{Wolf2011}%
  \BibitemOpen
  \bibfield  {author} {\bibinfo {author} {\bibfnamefont {J.}~\bibnamefont
  {Wolf}},\ }\href {http://www.ams.org/books/chel/372/} {\emph {\bibinfo
  {title} {Spaces of Constant Curvature}}}\ (\bibinfo  {publisher} {AMS Chelsea
  Publishing},\ \bibinfo {year} {2011})\BibitemShut {NoStop}%
\bibitem [{\citenamefont {Louko}\ and\ \citenamefont
  {Ruback}(1991)}]{Louko:1989up}%
  \BibitemOpen
  \bibfield  {author} {\bibinfo {author} {\bibfnamefont {J.}~\bibnamefont
  {Louko}}\ and\ \bibinfo {author} {\bibfnamefont {P.~J.}\ \bibnamefont
  {Ruback}},\ }\bibfield  {title} {\bibinfo {title} {{Spatially flat quantum
  cosmology}},\ }\href {https://doi.org/10.1088/0264-9381/8/1/013} {\bibfield
  {journal} {\bibinfo  {journal} {Classical and Quantum Gravity}\ }\textbf
  {\bibinfo {volume} {8}},\ \bibinfo {pages} {91} (\bibinfo {year}
  {1991})}\BibitemShut {NoStop}%
\bibitem [{\citenamefont {Allen}(1985)}]{Allen1985}%
  \BibitemOpen
  \bibfield  {author} {\bibinfo {author} {\bibfnamefont {B.}~\bibnamefont
  {Allen}},\ }\bibfield  {title} {\bibinfo {title} {Vacuum states in de
  {Sitter} space},\ }\href {https://doi.org/10.1103/PhysRevD.32.3136}
  {\bibfield  {journal} {\bibinfo  {journal} {Physical Review D}\ }\textbf
  {\bibinfo {volume} {32}},\ \bibinfo {pages} {3136} (\bibinfo {year}
  {1985})}\BibitemShut {NoStop}%
\bibitem [{\citenamefont {Elezovi\'c}(2013)}]{Ele}%
  \BibitemOpen
  \bibfield  {author} {\bibinfo {author} {\bibfnamefont {N.}~\bibnamefont
  {Elezovi\'c}},\ }\href@noop {} {\bibinfo {title} {Asymptotic expansions of
  exponentials of digamma function and identity for {Bernoulli} polynomials}}
  (\bibinfo {year} {2013}),\ \Eprint {https://arxiv.org/abs/1312.1604}
  {arXiv:1312.1604 [math.CA]} \BibitemShut {NoStop}%
\bibitem [{\citenamefont {Ferreira}\ and\ \citenamefont
  {López}(2004)}]{Ferreira2004}%
  \BibitemOpen
  \bibfield  {author} {\bibinfo {author} {\bibfnamefont {C.}~\bibnamefont
  {Ferreira}}\ and\ \bibinfo {author} {\bibfnamefont {J.~L.}\ \bibnamefont
  {López}},\ }\bibfield  {title} {\bibinfo {title} {Asymptotic expansions of
  the {Hurwitz}–{Lerch} zeta function},\ }\href
  {https://doi.org/https://doi.org/10.1016/j.jmaa.2004.05.040} {\bibfield
  {journal} {\bibinfo  {journal} {Journal of Mathematical Analysis and
  Applications}\ }\textbf {\bibinfo {volume} {298}},\ \bibinfo {pages} {210}
  (\bibinfo {year} {2004})}\BibitemShut {NoStop}%
\bibitem [{\citenamefont {Wong}(2001)}]{Wong2001}%
  \BibitemOpen
  \bibfield  {author} {\bibinfo {author} {\bibfnamefont {R.}~\bibnamefont
  {Wong}},\ }\href {https://books.google.co.uk/books?id=lwz2fBQTgB8C} {\emph
  {\bibinfo {title} {Asymptotic Approximations of Integrals}}},\ Classics in
  Applied Mathematics\ (\bibinfo  {publisher} {SIAM},\ \bibinfo {address}
  {Philadelphia},\ \bibinfo {year} {2001})\BibitemShut {NoStop}%
\end{thebibliography}%

\end{document}